\documentclass[journal, twoside]{IEEEtran}
\usepackage{pdfsync}

\usepackage{cite}
\usepackage{enumerate}
\usepackage{graphicx}
\usepackage[cmex10]{amsmath} % prevents amsmath from using a Type 3 font for math within footnotes
\usepackage{amssymb}
\usepackage{mathtools}
\usepackage{xspace}
\usepackage{subfigure}
\usepackage[lined,linesnumbered,ruled,commentsnumbered]{algorithm2e} % for algorithm
\usepackage{colonequals}
\usepackage[mathscr]{euscript}
\usepackage{fixltx2e}    % for text subscript
\usepackage{amsthm}
\usepackage{mathrsfs}
\usepackage[hidelinks]{hyperref}

\usepackage{stmaryrd} % for square brackets
\usepackage{xspace}

\usepackage{epsfig}
\usepackage{epstopdf}

\usepackage{tikz}
\usepackage{pgfplots}

\usetikzlibrary{arrows,shapes,backgrounds,plotmarks,positioning}

\pgfplotsset{compat=newest}                         % move axis labels close to the tick label automatically
\pgfplotsset{plot coordinates/math parser=false}
\newlength\figureheight
\newlength\figurewidth

\newtheorem{lemma}{Lemma}
\newtheorem{definition}{Definition}

\newtheorem{proposition}{Proposition}

\newtheorem{example}{Example}

\newcommand{\SEN}{\mathcal{S}}
\newcommand{\PUB}{\mathcal{X}}

\newcommand{\Sen}{S}
\newcommand{\Pub}{X}
\newcommand{\Rel}{\hat{X}}
\newcommand{\sen}{s}
\newcommand{\pub}{x}
\newcommand{\rel}{\hat{x}}

\newcommand{\Edge}{\mathcal{E}}
\newcommand{\G}{\mathcal{G}}
\newcommand{\GG}{\mathbb{G}}
\newcommand{\TGG}{\tilde{\mathbb{G}}}
\newcommand{\E}{\mathbb{E}}

\newcommand{\Set}[1]{\{#1\}}
\newcommand{\range}[1]{\llbracket{#1}\rrbracket}

\newcommand{\GK}{G\'{a}cs-K\"{o}rner\ }

\newcommand{\Pat}{\mathcal{P}}
\newcommand{\Lat}{\mathcal{L}}
\newcommand{\Qat}{\mathcal{Q}}

\newcommand{\Link}{\leftrightsquigarrow}

\DeclareMathOperator*{\argmax}{arg\,max}
\DeclareMathOperator*{\argmin}{arg\,min}

\begin{document}

\title{Developing Non-Stochastic Privacy-Preserving Policies Using Agglomerative Clustering}

\author{Ni Ding,~\IEEEmembership{Member,~IEEE,} and Farhad Farokhi,~\IEEEmembership{Senior Member,~IEEE}\thanks{The authors are with the University of Melbourne.}\thanks{emails:\{ni.ding,farhad.farokhi\}@unimelb.edu.au}
\thanks{The work of Ni Ding is funded by the Doreen Thomas Postdoctoral Fellowship at the University of Melbourne.}
\thanks{The work of F. Farokhi is funded by the Melbourne School of Engineering at the University of Melbourne.}
}

\markboth{IEEE journal}
{Ding \MakeLowercase{\textit{et al.}}: Developing Non-Stochastic Privacy-Preserving Policies Using Agglomerative Clustering}



\maketitle

\begin{abstract}
We consider a non-stochastic privacy-preserving problem in which an adversary aims to infer sensitive information $\Sen$ from publicly accessible data $\Pub$ without using statistics.
We consider the problem of
generating and releasing a quantization $\Rel$ of $\Pub$ to minimize the privacy leakage of $\Sen$ to $\Rel$ while maintaining a certain level of utility (or, inversely, the quantization loss).
The variables $\Sen$ and $\Pub$ are treated as bounded and non-probabilistic, but are otherwise general. We consider two existing non-stochastic privacy  measures, namely the maximum uncertainty reduction $L_0(\Sen \rightarrow \Rel)$ and the refined information $I_*(\Sen; \Rel)$ (also called the maximin information) of $\Sen$.
For each privacy measure, we propose a corresponding agglomerative clustering algorithm that converges to a locally optimal quantization solution $\Rel$ by iteratively merging elements in the alphabet of $\Pub$.
To instantiate the solution to this problem, we consider two specific utility measures, the worst-case resolution of $\Pub$ by observing $\Rel$ and the maximal distortion of the released data $\Rel$.
We show that the value of the maximin information $I_*(\Sen; \Rel)$ can be determined by dividing the confusability graph into connected subgraphs. Hence, $I_*(\Sen; \Rel)$ can be reduced by merging nodes connecting subgraphs.
The relation to the probabilistic information-theoretic privacy is also studied by noting that the G{\'a}cs-K{\"o}rner common information is the stochastic version of $I_*$ and indicates the attainability of statistical indistinguishability.
\end{abstract}


% \fi
\IEEEpeerreviewmaketitle

\section{Introduction}
\IEEEPARstart{N}{owadays}
we share and exchange data with others regularly while being increasingly concerned about whether our personal data is well protected.
In particular, in this big data era, advances in efficient data analytics have improved an adversary's ability in obtaining individuals' private information \cite{Lane2014BigData}.
For example, machine learning models with high accuracy can potentially reveal individual labels of the training data as in the case of membership information attacks \cite{Li2013Membership}.
In these cases, the privacy attack can happen during the legitimate use of data, e.g., querying the dataset, survey report, and speech recognition, without  direct access to  confidential data.
Thus, the meaning of privacy has moved far beyond its original definition in~\cite{Glancy1979RightPriv} and cannot be achieved by just `anonymizing' or `secluding' the sensitive data from release.
%and at the same time should take into consideration the quality of the anonymized or sanitized data.

For a user in the public domain who is interested in the aggregated statistics and interacts with a data curator using queries, the differential privacy (DP) \cite{Dwork2011DP} provides a mathematical definition for the privacy loss: an upper bound on how the statistics of sanitized/randomized responses to a query change with and without an individual's record.
By replacing query response with data mining, e.g., maximum \textit{a posteriori} estimate, DP can be tuned to a privacy measure in statistical inference problems, e.g., in machine learning \cite{Chaudhuri2009DPLogisticReg,Friedman2010DPDataMine}.
While DP is concerned with the individual indistinguishability of the released data, privacy in information theory \cite{PvsInfer2012,Asoodeh2014Note,Liao2019Alpha} refers to the exact amount (e.g., the number of bits) of sensitive information that is leaked to the public. This enables study of privacy-utility tradeoff (PUT) via the problem of minimizing the privacy leakage subject to a constraint on the utility/usefulness of the released data.

In statistical data mining, unknown quantities are conventionally treated as random variables, measurable functions that map the probability space to the event space.
As a result, privacy is usually measured as a stochastic feature using, e.g., the mutual information in \cite{PvsInfer2012} or the R\'{e}nyi divergence\footnote{DP can be defined as the supremum of the R\'{e}nyi divergence $D_{\alpha}(p(\cdot) \| q(\cdot))$ for $\alpha = \infty$ \cite{Dwork2018DPPredict}.} in \cite{Liao2019Alpha,Dwork2011DP}.
However, in some practical applications, it may be preferred to measure and optimize privacy or information leakage in a non-stochastic setting.
For instance, consider the problem of releasing a table of just 50 records. The size is not large enough for making probabilistic inference about the population\footnote{This could be the case in a small experiment or at the beginning of data release in a gradual data release setting. For example, in small datasets, each row of a tabular dataset is often distinct and the information only depends on the range, the number of distinct values, and not probabilities or statistics.
We show an example of this situation in Section~\ref{sec:Hungarian}.}.
The data curator's concern could be whether an adversary can gain private information that is not based on statistics.
Kolmogorov had partly answered this question in his early study \cite{Kolmogorov1959Epsilon} stating that the more values of sensitive data appears concurrently with the released data, the less certain an adversary is about the sensitive data.
This is also an information-theoretic interpretation of the $k$-anonymity \cite{KAnony2002}
and motivated the adoption of non-stochastic privacy metrics~\cite{Farhad2018,Farokhi2019HT}.
Non-stochastic treatment of variables and measures of information have proved popular in other engineering domains, such as networked control and estimation in which variables might be bounded but do not follow a well-defined probability distribution \cite{Nair2013,Saberi2019StateEst,duan2014transfer,wiese2016uncertain}.

In this paper, we study how to preserve privacy against non-stochastic inference, where
the adversary is able to deduce non-statistical features of the data while both privacy and utility are measured by non-probabilistic information quantities.
Specifically, for a data curator who wants to share non-sensitive data $\Pub$ with the public, but to protect the sensitive data $\Sen$ that is related to $\Pub$, the problem can be cast as generating a sanitization $\Rel$ with a specified level of the data fidelity/utility while leaking the least amount of information about $\Sen$.
Based on non-stochastic information theory, two privacy measures are considered: $L_0(\Sen \rightarrow \Rel)$ capturing the information on $\Sen$ conveyed by $\Rel$~\cite{Farhad2018} and the \emph{maximin information} $I_*(\Sen ; \Rel)$ measuring the adversary's knowledge on $\Sen$ refined by $\Rel$~\cite{Nair2013}.
We reveal different interpretations in privacy between these two measures: $L_0(\Sen \rightarrow \Rel)$ measures the maximum uncertainty reduction on $\Sen$ at the adversary side, which is shown to correspond to $k$-anonymity, %\footnote{The definition of $L_0$ is based on the no-stochastic information measure in \cite{Kolmogorov1959Epsilon}.}
while $I_*(\Sen; \Rel)$ measures how distinguishable $\Sen$ can be by observing $\Rel$, or how much private information can be obtained without error by the adversary.

We consider the privacy-preserving problem by generating $\Rel$ that minimizes the privacy leakage, $L_0(\Sen \rightarrow  \Rel)$ or $I_*(\Sen; \Rel)$ subject to maintaining the utility $U(\Pub ; \Rel)$ of the released data above some threshold.
The non-stochastic sanitization $\Rel$ is done by a quantization method, or a deterministic clustering of the alphabet of the public data $\Pub$. To this end,
we propose a greedy clustering algorithm for extracting a locally optimal solution $\Rel$ by iteratively merging two elements of $\Pub$ that strictly reduces the Lagrangian function of the privacy-preserving problem.
%$L_0(\Sen \rightarrow \Pub) - \lambda U(\Pub; \Rel)$ or $I_*(\Sen ; \Pub) - \lambda U(\Pub; \Rel)$
%
We show that the value of the maximin information $I_*(\Sen; \Pub)$ is determined by the maximum number of disconnected subgraphs in an undirected uncapacitated graph, which is equivalent to the \emph{confusability graph} for the study on Shannon's zero-error capacity~\cite{Erickson2013Book}.
Therefore, the proposed agglomerative clustering algorithm for minimizing $I_*(\Sen; \Rel)$ is analogous to a subgraph merging process.
To instantiate the solution to the problem, we study two specific utility measures for $U(\Pub;\Rel)$: the minimum uncertainty reduction $I_0(\Pub\rightarrow\Rel)$, the worst-case resolution of the public data $\Pub$ via the released data $\Rel$~\cite{Nair2013}, and the maximum distortion $\max_{\pub,\rel} d(\pub,\rel)$ of the sanitized data $\Rel$.
We also investigate the relationship between our work and the stochastic information-theoretic privacy studies in \cite{PvsInfer2012,Asoodeh2014Note,Liao2019Alpha}. We elaborate that the G{\'a}cs-K{\"o}rner common information \cite{GacsKorner1973} is the stochastic version of the maximin information, indicating when statistical indistinguishability, e.g., the DP \cite{Dwork2011DP}, is attainable.

Finally, we show how to apply the proposed agglomerative clustering algorithm to plot the Pareto frontier, where one can directly search for a privacy-preserving data sanitization solution for any given utility constraint. We also run experiments on a real-world dataset to show that the proposed agglomerative clustering algorithm is capable of realistic/practical privacy-preserving sanitization.

\textbf{Related works}:
For a continuous sensitive variable $\Sen$, the problem of designing the quantization $f(\Sen)$ and its sanitization $\hat{f}(\Sen)$ to minimize the privacy leakage $L_0(\Sen \rightarrow \hat{f}(\Sen))$ or $I_*(\Sen ; \hat{f}(\Sen))$ subject to an upper bound on the $\ell_2$-norm $\|f(\Sen) - \hat{f}(\Sen)\|_2$ was investigated in~\cite{Farhad2018}.
This reduces to the privacy preserving problem in this paper when $\Pub = f(\Sen)$. This paper generalizes the framework of \cite{Farhad2018} as there is no requirement that $\Pub$ is a deterministic function of $\Sen$.
Moreover, the difference between $L_0(\Sen \rightarrow \hat{f}(\Sen))$ and $I_*(\Sen ; \hat{f}(\Sen))$, the graph decomposition method for determining $I_*(\Sen ; f(\Sen))$ and its role in minimizing $I_*(\Sen ; \hat{f}(\Sen))$ are not discussed in \cite{Farhad2018}.
The setup of this paper matches the typical setting in statistical inference\footnote{Statistical inference considers the situation where a user/adversary is planning to deduce properties $\Sen$ of a population $\Pub$ based on noisy observations $\Rel$, where the statistics of the population $\Pub$ is usually given or specified.}
and is adopted in most of the stochastic information-theoretic privacy studies \cite{PvsInfer2012,Asoodeh2014Note,Liao2019Alpha} (see Section~\ref{sec:ToITPriv}).

\textbf{Organization}: This paper is organized as follows. Section~\ref{sec:Preliminary} reviews some definitions in non-stochastic information theory. Section~\ref{sec:PrivMeasure} introduces privacy measures $L_0$ and $I_*$. Section~\ref{sec:GraphDecomp} proposes a graph decomposition method for determining $I_*$. In Section~\ref{sec:Algo}, we formulate the non-stochastic privacy-preserving problem and propose two agglomerative clustering algorithms. Section~\ref{sec:ToITPriv} studies the relationship between our work and the stochastic information-theoretic privacy. Section~\ref{sec:Exp} presents the experimental results.

\section{Non-stochastic Information Theory}
\label{sec:Preliminary}

In non-stochastic information theory \cite{Nair2013}, for the sample space $\Omega$, we say that the mapping $X:\Omega \rightarrow \mathbb{X}$ is an \emph{uncertain variable (uv)}. For a sample $\omega \in \Omega$, $X(\omega)$ denote a realization of the uv $X$. The notation $X(\omega)$ is simplified to $X$ in the rest of the paper if the context is clear.
When $\mathbb{X}$ is a finite set, $X$ is a \emph{discrete uv}. In this paper, we only consider discrete uvs.
Here, a uv differs from the a random variable (rv) in probability theory in that we do not assume a $\sigma$-algebra subset family over the sample space $\Omega$ or a probability measure over this subset family.
For a pair of discrete uvs $\Sen$ and $\Pub$, let $\range{\Sen, \Pub } \triangleq \Set{(\Sen(\omega),\Pub(\omega)) \colon \omega \in \Omega}$ and $\range{\Sen | x} \triangleq \Set{\Sen(\omega) \colon \Pub(\omega) = \pub, \omega \in \Omega}$, respectively, be the joint range of $\Sen$ and $\Pub$ and the conditional range of $\Sen$ based on the observation that $\Pub=\pub$. Note, we can rewrite $\range{\Sen, \Pub } = \bigcup_{\pub \in \range{\Pub}} \big( \range{\Sen | \pub} \times \Set{\pub} \big)$ and $\range{\Sen | \Pub} = \Set{\range{\Sen | \pub} \colon \pub \in \range{\Pub}}$.
We say that $\Sen$ and $\Pub$ are \emph{independent}, if
$ \range{\Sen | \pub} = \range{\Sen},  \forall \pub \in \range{\Pub}. $
For $\range{\Sen,\Pub}$, we have the marginal ranges $\range{\Sen}$ and $\range{\Pub}$.
The non-probabilistic entropy of a uv $\Pub$ is defined as
$$ H_0 (\Pub) \triangleq \log |\range{\Pub}|, $$
which coincides with the Hartley (maximum) entropy \cite{Hartley1928TransInf} or the R\'{e}nyi entropy $H_{\alpha}$ in the case $\alpha = 0$ \cite{Renyi1961Measure}. We assume the base of all logarithms is $2$ in this paper.

\section{Non-stochastic Inference and Leakage}
\label{sec:PrivMeasure}

The sensitive/private data is denoted by $\Sen$. Consider the case where a data curator wants to release $\Pub$ to the public, which is related\footnote{Relatedness, similar to correlation in the probabilistic sense, states that $\range{\Sen | \pub} = \range{\Sen}$ does not hold for all $\pub \in \range{\Pub}$, or $\Sen$ and $\Pub$ are not non-stochastically independent uvs.} with $\Sen$.
For example, the taxation office may want to release the income records, which, even if anonymized,  could be observed by an adversary in the public domain to infer individuals' identities.
The data curator needs to maintain a certain level of usefulness/utility of the released data, e.g., to the ensure correctness of answers to legitimate surveys/queries. At the same time, the curator has privacy concerns in the sense that the more fidelity the released data has, the easier an adversary can infer the private data $\Sen$.
Thus, it is necessary for the curator to be equipped with a valid privacy measure to anticipate the risk of the information leakage.

\subsection{Uncertainty Reduction}

For any realization $\pub$ of the public data $\Pub$, Kolmogorov defined a `combinatorial' conditional entropy as $\log{\range{\Sen | \pub}}$ in \cite{Kolmogorov1959Epsilon}.
This interprets the uncertainty/entropy reduction $\log (|\range{\Sen}|/|\range{\Sen | \pub}|)$ as a measure of non-stochastic information on $\Sen$ gained by an adversary in the public domain after observing $\pub$, or the privacy leakage at the data curator side by releasing $\pub$.
For the conditional information in the sense of the maximum entropy $ H_0(\Sen | \Pub) \triangleq \max_{\pub \in \range{\Pub}} \log |\range{\Sen | \pub}|$, the author in \cite{Nair2013} proposed the
\emph{non-stochastic $0$-information} as the minimal difference between prior and posterior entropy
$$ I_0(\Sen \rightarrow \Pub) \triangleq H_0(\Sen) - H_0(\Sen | \Pub) = \min_{\pub \in \range{\Pub}} \log \frac{|\range{\Sen}|}{ |\range{\Sen | \pub}|}. $$
However, in \cite{Farhad2018}, it was suggested to consider the worst-case, i.e., the minimal posterior entropy/uncertainty $B_0(\Sen | \Pub) \triangleq \min_{\pub \in \range{\Pub}} \log |\range{\Sen | \pub}|$, to quantify the \emph{maximal information leakage}
$$ L_0(\Sen \rightarrow \Pub) \triangleq H_0(\Sen) - B_0(\Sen | \Pub) = \max_{\pub \in \range{\Pub}} \log \frac{|\range{\Sen}|}{ |\range{\Sen | \pub}|}. $$
as a measure of privacy.

\subsection{Maximin Information/Indistinguishability}

Note that neither $I_0$ or $L_0$ is symmetric. That is, $I_0(\Sen \rightarrow \Pub) = I_0(\Pub \rightarrow \Sen)$ or $L_0(\Sen \rightarrow \Pub) = L_0(\Pub \rightarrow \Sen)$ do not hold in general. However, a symmetric measure of the dependence between $\Sen$ and $\Pub$ can be defined based on the concept of overlap partition.
\begin{definition}[Overlap Partition {\cite[Definitions~3.1 and 3.2, Lemma~3.1]{Nair2013}}]\hfill\break\vspace{-1em} \label{def:OverPart}
    \begin{enumerate}[(a)]
		\item \textbf{overlap connectedness}: A pair $\sen,\sen' \in \range{\Sen}$ is $\range{\Sen | \Pub}$-overlap connected, denoted by $ \sen \Link \sen'$, if there exists an ordered finite sequence $(\pub_1,\dotsc, \pub_n)\in \range{\Pub}^n$ such that $\sen \in \range{\Sen | \pub_1}$, $\sen' \in \range{\Sen | \pub_n}$ and $\range{\Sen | \pub_i } \cap \range{\Sen | \pub_{i+1}} \neq \emptyset$ for all $i \in \Set{1,\dotsc,n-1}$. Thus, $\sen \Link \sen, \forall \sen \in \range{\Sen}$ and $\sen \Link \sen', \forall \sen,\sen' \in \range{\Sen | \pub}$, $x \in \range{\Pub}$. A subset $\SEN \subseteq \range{\Sen}$ is $\range{\Sen | \Pub}$-overlap connected if $\sen \Link \sen' , \forall \sen, \sen' \in \SEN$;
        \item \textbf{overlap isolation}: Two subsets $\SEN, \SEN' \subseteq \range{\Sen}$ are $\range{\Sen | \Pub}$-overlap isolated if there does not exist $\sen \in \SEN$ and $\sen' \in \SEN'$ such that $ \sen \Link \sen'$;
        \item \textbf{overlap isolated partition}: A partition of $\range{\Sen}$ is a $\range{\Sen | \Pub}$-overlap isolated partition, denoted by $\Pat_{\range{\Sen|\Pub}}$, if every distinct $\SEN, \SEN' \in \Pat_{\range{\Sen|\Pub}}$ are $\range{\Sen | \Pub}$-overlap isolated.
               \item \textbf{overlap partition}: A $\range{\Sen | \Pub}$-overlap isolated partition $\Pat_{\range{\Sen|\Pub}}$ is called $\range{\Sen | \Pub}$-overlap partition if each $\SEN \in \Pat_{\range{\Sen|\Pub}}$ is $\range{\Sen | \Pub}$-overlap connected;
        \item \textbf{maximin information}: Let $\Pat_{\range{\Sen|\Pub}}^*$ be the finest $\range{\Sen | \Pub}$-overlap partition with the largest size $|\Pat_{\range{\Sen|\Pub}}^*|$. Such partition is unique. The maximin information is defined as
            $$ I_*(\Sen;\Pub) = \log|\Pat_{\range{\Sen|\Pub}}^*|. $$
    \end{enumerate}
\end{definition}

The maximin information $I_*(\Sen; \Pub)$ measures the `refined' knowledge \cite[Section~III-B]{Nair2013}, or the highest resolution, on the range $\range{\Sen}$ by observing $\Pub$.
We show in the next subsection that $I_*(\Sen ; \Pub)$ indicates the non-stochastic distinguishability in the view of privacy.

\subsection{$L_0(\Sen \rightarrow \Pub)$, $I_*(\Sen;\Pub)$, $k$-anonymity and Non-stochastic Distinguishability }
\label{subsec:PrivRelation}

For the release of tabular data, \emph{$k$-anonymity} was proposed in \cite{KAnony2002} to guarantee that an adversary cannot distinguish between at least $k$ rows of the record for each instance of the released data.
\begin{definition}[$k$-anonymity] \label{def:Kanony}
    $\range{\Sen | \Pub}$ is $k$-anonymous if $|\range{\Sen | \pub}| \geq k$ for all $\pub \in \range{\Pub}$.
\end{definition}
The method for attaining $k$-anonymity is to ensure $k$-occurrences of the records $|\range{\Sen | \pub}| = k$ for each $\pub$ in the released table.\footnote{The value $k$ in the original definition \cite[Definition~3]{KAnony2002} refers to the number of records, including $k$ records with the same value of the sensitive data $\sen$ such that $\sen \in \range{\Sen | \pub}$. It is shown in \cite{LDiv2006,Farhad2018} that a $k$-anonymous database with the identical occurrences of the sensitive variable is subject to the homogeneity attack \cite{LDiv2006,Farhad2018}. This does not apply to the $k$-anonymity in Definition~\ref{def:Kanony}, which is a more strict definition where $k$ refers to the number of distinct values of $\Sen$.}
A straightforward result from Definition~\ref{def:Kanony} is the one-to-one correspondence between $L_0$ and the $k$-anonymity.

\begin{lemma}[$L_0 \Longleftrightarrow k$-anonymity] \label{lemma2}
    $\range{\Sen | \Pub}$ is $k$-anonymous if and only if $B_0(\Sen | \Pub) \geq \log k$ or $L_0(\Sen \rightarrow \Pub) \leq \log (|\range{\Sen}|/k)$. \hfill \IEEEQED
\end{lemma}

While $L_0$ or the $k$-anonymity denotes the non-stochastic uncertainty reduction, $I_*(\Sen; \Pub)$ measures the non-stochastic distinguishability.
In \cite{Farokhi2019HT}, the non-stochastic privacy is measured as the change of the range of the public data $|\range{\Pub | \sen} \setminus \range{\Pub | \sen'} \sqcup \range{\Pub | \sen'} \setminus \range{\Pub | \sen} |$ conditioned on two distinct sensitive uv values $\sen$ and $\sen'$.
Here, $|\range{\Pub | \sen} \setminus \range{\Pub | \sen'} \sqcup \range{\Pub | \sen'} \setminus \range{\Pub | \sen} |$ measures the distinguishability between $\sen$ and $\sen'$ (the lower the value, the less distinguishable and the more privacy).
This can be viewed as a non-stochastic version of the differential privacy (DP) \cite{Dwork2011DP}.
Clearly, if $I_*(\Sen ; \Pub) > 0 $, there are at least two non-overlapping/disjoint $\SEN,\SEN' \subsetneq \Pat_{\range{\Sen | \Pub}}^*$ such that an adversary can discriminate perfectly any pair of $\sen \in \SEN$ and $\sen' \in \SEN'$ by observing $\Pub$.\footnote{This is because $|\range{\Pub | \sen} \setminus \range{\Pub | \sen'} \sqcup \range{\Pub | \sen'} \setminus \range{\Pub | \sen} |$ attains the maximum if $\range{\Pub | \sen} \setminus \range{\Pub | \sen'} = \range{\Pub | \sen}$ and $\range{\Pub | \sen'} \setminus \range{\Pub | \sen} = \range{\Pub | \sen'}$.}
In this case, any distinguishability-based privacy measure is minimized (no privacy).
It also relates to the zero-error capacity \cite{ZeroErrorC1956}: $I_*(\Sen; \Pub) $  is the largest information amount (in bits) of $\Sen$ that can be conveyed without any error by transmitting $\Pub$ \cite[Theorem~4.1]{Nair2013}.
In addition, we can prove the following relationship between $I_*$ and $L_0$. The proof is presented in Appendix~\ref{proof:prop:1}.

\begin{proposition} \label{prop:1}
    $I_* (\Sen ; \Pub) \leq L_0(\Sen \rightarrow \Pub)$. \hfill \IEEEQED
\end{proposition}

This means that $L_0(\Sen \rightarrow \Pub)$ could be nonzero when $0 = I_* (\Sen ; \Pub)$, i.e., even if the adversary cannot perfectly obtain some bits of $\Sen$ from $\Pub$, he/she might still become more certain about $\Sen$.
%
%Using Proposition~\ref{prop:1}, we can interpret the case $0 = I_* (\Sen ; \Pub) \leq L_0(\Sen \rightarrow \Pub)$ as even if the adversary cannot obtain some bits of $\Sen$ perfectly from $\Pub$, he/she might still become more certain on $\Sen$.
%%
Apart from these non-stochastic measures, we show in Section~\ref{sec:ToITPriv} how $L_0$ and $I_*$ relate to stochastic information theory.

\section{Graph Decomposition for Determining $I_*(\Sen;\Pub)$}
\label{sec:GraphDecomp}

Let $G_{\Pub} = (\range{\Pub},\Edge_{\range{\Sen | \Pub}})$ be an undirected uncapacitated graph with the node set $\range{\Pub}$ and the edge set $\Edge_{\range{\Sen | \Pub}} = \Set{(\pub,\pub') \colon  \range{\Sen | \Pub = \pub} \cap \range{\Sen | \Pub = \pub'} \neq \emptyset}$.
Here, $G_{\Pub}$ corresponds to the \emph{adjacency matrix} in stochastic information theory~\cite{ZeroErrorC1956} $A = [a_{\pub,\pub'}]$ where $a_{\pub,\pub'} = 1$ if $\exists \sen \in \range{\Sen}$ such that the joint probabilities $p(\sen,\pub)$ and $p(\sen,\pub')$ are both non-zero and $a_{\pub,\pub'} = 0$ otherwise.
Therefore, $G_{\Pub}$ is also called the \emph{confusability graph}~\cite{Erickson2013Book}.\footnote{The adjacency matrix $A$ in~\cite{ZeroErrorC1956} and the corresponding confusability graph in~\cite{Erickson2013Book} only depend on whether the joint probability $p(\sen,\pub)$ is nonzero or not. By knowing that each non-zero $p(\sen,\pub)$ determines the presence of $(\sen,\pub)$ in $\range{\Sen,\Pub}$, the stochastic joint probability $p(\sen,\pub), \forall \sen,\pub$ can be reduced to the non-stochastic joint range by $\range{\Sen,\Pub} = \Set{(\sen,\pub) \colon p(\sen,\pub) \neq 0}$. This explains that both the Shannon capacity of a graph, the zero-error capacity, and the maximin information $I_*(\Sen,\Pub)$ are related to the graph theory or combinatorics~\cite{Lovasz1979,ZeroErrorInfo1998}. }

\begin{figure}[tpb]
	\centering
    \scalebox{0.6}{\begin{tikzpicture}
\begin{axis}[
width=3.8in,
height=1.6in,
scale only axis,
xmin=0.5,
xmax=7.5,
xtick = {1,2,3,4,5,6,7},
xticklabels = {$\pub_1$,$\pub_2$,$\pub_3$,$\pub_4$,$\pub_5$,$\pub_6$,$\pub_7$},
xlabel={\large public data $\Pub$},
ymin=0.5,
ymax=6.5,
ytick = {1,2,3,4,5,6},
yticklabels = {$\sen_1$,$\sen_2$,$\sen_3$,$\sen_4$,$\sen_5$,$\sen_6$},
ylabel={\large sensitive data $\Sen$},
scatter/classes={
a={mark=square,blue},%
b={mark=square*,blue, mark size =3}},
grid=major]
% \addplot[] is better than \addplot+[] here:
% it avoids scalings of the cycle list
\addplot[scatter,only marks,
scatter src=explicit symbolic]
coordinates {
(1,1) [b]
(2,1) [b]
(1,2) [b]
(3,3) [b]
(4,3) [b]
(5,4) [b]
(6,5) [b]
(7,6) [b]
};
\end{axis}
\end{tikzpicture} }
	\caption{The joint range $\range{\Sen,\Pub}$ of uvs $\Sen$ and $\Pub$ with $\range{\Sen} = \Set{\sen_1,\dotsc,\sen_6}$ and $\range{\Pub} = \Set{\pub_1,\dotsc, \pub_7}$. A blue square at $(\sen_i,\pub_j)$ denotes $(\sen_i,\pub_j) \in \range{\Sen,\Pub}$; otherwise $(\sen_i,\pub_j) \notin \range{\Sen,\Pub}$.}
	\label{fig:Psx}
\end{figure}

A \textit{decomposition} $\Pat_{G_{\Pub}}$ of $G_{\Pub}$ is a partition of $\range{\Pub}$ such that any two distinct subgraphs $\mathcal{\Pub}, \mathcal{\Pub}' \in \Pat_{G_{\Pub}}$ are disconnected; The \textit{finest decomposition}, denoted by $\Pat_{G_{\Pub}}^*$, is a decomposition such that each subgraph $\PUB$ is connected \cite{Cunningham1985NetStrength}.
The following lemma states the equivalence between graph decomposition and the $\range{\Sen | \Pub}$-overlap partition. The proof is in Appendix~\ref{app:lemma:Graph}.
\begin{lemma} \label{lemma:Graph}
     For each decomposition $\Pat_{G_{\Pub}}$, $\Set{\range{\Sen | \PUB} : \mathcal{\Pub} \in \Pat_{G_{\Pub}}}$ is a $\range{\Sen | \Pub}$-overlap isolated partition; For the finest decomposition $\Pat_{G_{\Pub}}^*$, $\Pat_{\range{\Sen | \Pub}}^* = \Set{\range{\Sen | \PUB} : \PUB \in \Pat_{G_{\Pub}}^*} $ is the unique $\range{\Sen | \Pub}$-overlap partition so that the maximin information is
      $ I_*[\Sen ; \Pub] = \log|\Pat_{G_{\Pub}}^*|$. \hfill \IEEEQED
\end{lemma}
 For $\mathcal{\Pub} \subseteq \range{\Pub}$, let $\kappa(\mathcal{\Pub}) = |\{(x,x') \in \Edge_{\range{\Sen | \Pub}}: x \in \mathcal \Pub, x'\in \range{\Pub} \setminus \mathcal{\Pub}\}|$ denote the value of the cut $\Set{\mathcal{\Pub},\range{\Pub} \setminus \mathcal{\Pub}}$. The graph $G_{\Pub}$ is decomposable if the min-cut
$ \min \Set{\kappa(\mathcal{\Pub}): \emptyset \neq \mathcal{\Pub} \subsetneq \range{\Pub}} = 0$.
The solution to the min-cut problem $ \argmin \Set{\kappa(\mathcal{\Pub}): \emptyset \neq \mathcal{\Pub} \subsetneq \range{\Pub}} $ forms a set lattice \cite[Section~2.3]{Fujishige2005}\footnote{More precisely, the set $\mathcal{T} = \argmin \Set{\kappa(\mathcal{\Pub}): \emptyset \neq \mathcal{\Pub} \subsetneq \range{\Pub}} \sqcup \Set{\emptyset} \sqcup \Set{\range{\Pub}}$ is a lattice such that, for any $\PUB_1,\PUB_2 \in \mathcal{T}$, $\PUB_1 \cap \PUB_2 \in \mathcal{T}$ and $\PUB_1 \cup \PUB_2 \in \mathcal{T}$. } and can be determined by the max-flow algorithm \cite{MaxFlow1988} in polynomial time.
All the smallest min-cut solutions constitute the finest decomposition \cite[Section~2.2]{Fujishige2005}:
$$ \Pat_{G_{\Pub}}^* = \bigcap \argmin \Set{\kappa(\mathcal{\Pub}): \emptyset \neq \mathcal{\Pub} \subsetneq \range{\Pub}}.$$
A simple method to determine $\Pat_{G_{\Pub}}^*$ is to recursively run either breadth-first search (BFS) or death-first search (DFS) \cite{BFSDFS1973}. We run BFS or DFS staring from any node $\pub \in \Pub$ to find all other nodes that are reachable (via some path) from $\pub$. This identifies a connected subgraph in $\Pat_{G_{\Pub}}^*$. Then, we repeat this procedure for any node that has not been searched until no such node is left.
The complexity of BFS/DFS is $O(|\range{\Pub}| + |\Edge_{\range{\Sen | \Pub}}|)$. The above method calls BFS/DFS for $|\Pat_{G_{\Pub}}^*|$ times and therefore the complexity is upper bounded by $O(|\range{\Pub}|^2 + |\range{\Pub}||\Edge_{\range{\Sen | \Pub}}|)$.

There are also other methods that determine $\Pat_{G_{\Pub}}^*$. For example, one of algorithms in \cite{Cunningham1985NetStrength,Chen1994NetStrenght,Kolmogorov2010NetPSP} for determining the network strength.\footnote{If the network strength is nonzero, $G_{\Pub}$ is connected and $I(\Sen; \Pub) = 0$. However, if the network strength is zero, $G_{\Pub}$ is decomposable and the algorithms in \cite{Cunningham1985NetStrength,Chen1994NetStrenght,Kolmogorov2010NetPSP} return the finest decomposition $\Pat_{G_{\Pub}}^*$ \cite{Cunningham1985NetStrength}.}
The lowest complexity of these algorithms is $O(|\range{\Pub}|)$ runs of the max-flow algorithm.
Without explicitly constructing the graph $G_{\Pub}$, we can still obtain the Laplacian of the confusability matrix to determine the finest decomposition $\Pat_{G_{\Pub}}^*$; see \cite[Remark~6]{GacsKorner2016arXiv}. Note that the validity of this Laplacian method is due to the inherit relationship between the maximin information $I_*$ and \GK common information, which is explained in Section~\ref{sec:ToITPriv}.

\begin{figure}[tpb]
	\centering
    \scalebox{0.8}{\begin{tikzpicture}

%nodes
\draw (0.5,1) circle (0.3);
\node at (0.5,1) {\Large $\pub_2$};

\draw (0.5,-1) circle (0.3);
\node at (0.5,-1) {\Large $\pub_4$};

\draw (-1.5,1) circle (0.3);
\node at (-1.5,1) {\Large $\pub_1$};

\draw (-1.5,-1) circle (0.3);
\node at (-1.5,-1) {\Large $\pub_3$};

\draw (2,0.7) circle (0.3);
\node at (2,0.7) {\Large $\pub_5$};

\draw (2,-0.7) circle (0.3);
\node at (2,-0.7) {\Large $\pub_6$};

\draw (3.5,0) circle (0.3);
\node at (3.5,0) {\Large $\pub_7$};

% edges
\draw (-1.2,1) --  (0.2,1);
\draw (0.2,-1) -- (-1.2,-1);

\end{tikzpicture} }
	\caption{The undirected graph $G_{\Pub} = (\range{\Pub}, \Edge_{\range{\Sen | \Pub}})$ based on the conditional range $\range{\Sen | \Pub}$ of Fig.~\ref{fig:Psx}. Both the min-cut and graph connectivity of $G_{\Pub}$ is $0$, i.e., $G_{\Pub}$ is disconnected, which means $I_*(\Sen ; \Pub) > 0$. The finest decomposition of $G_{\Pub}$ is $\Pat_{G_{\Pub}}^* = \Set{\Set{\pub_1,\pub_2},\Set{\pub_3,\pub_4},\Set{\pub_5},\Set{\pub_6},\Set{\pub_7}}$, which results in $I_*(\Sen ; \Pub) = \log |\Pat_{G_{\Pub}}^*| = \log 5$.}
	\label{fig:Gx}
\end{figure}

\begin{example} \label{ex:main}
    For $\range{\Sen} = \Set{\sen_1,\dotsc,\sen_6}$, $\range{\Pub} = \Set{\pub_1,\dotsc, \pub_7}$ and the joint range $\range{\Sen,\Pub}$ shown in Fig.~\ref{fig:Psx}, we have $G_{\Pub} = (\range{\Pub}, \Edge_{\range{\Sen | \Pub}} )$ in Fig.~\ref{fig:Gx} with the min-cut being $0$, i.e., the graph is decomposable. The finest decomposition $\Pat_{G_X}^* = \Set{\Set{\pub_1,\pub_2},\Set{\pub_3,\pub_4},\Set{\pub_5},\Set{\pub_6},\Set{\pub_7}}$ and the set of all min-cut solutions can be constructed by the fusion of subsets in $\Pat_{G_X}^*$:
    \begin{multline}
        \argmin \Set{\kappa(\mathcal{\Pub}): \emptyset \neq \mathcal{\Pub} \subsetneq \range{\Pub}} =  \\
                \Big\{ \bigsqcup_{\PUB \in \mathfrak{X}} \PUB  \colon  \mathfrak{X} \subseteq \Pat_{G_X}^*\Big\} \setminus \Set{\range{\Pub}}. \nonumber
    \end{multline}
    For example, $\Set{\pub_1,\dotsc, \pub_4} =  \Set{\pub_1,\pub_2} \sqcup \Set{\pub_3,\pub_4}$ is a min-cut solution and so is $\Set{\pub_5,\pub_6,\pub_7} = \Set{\pub_5} \sqcup \Set{\pub_6} \sqcup \Set{\pub_7}$. The finest decomposition $\Pat_{G_{\Pub}}^*$ determines the unique $\range{\Sen | \Pub}$-overlap partition $\Pat_{\range{\Sen | \Pub}}^* = \Set{\range{\Sen | \PUB} \colon \PUB \in \Pat_{G_{\Pub}}^*} = \Set{\Set{\sen_1,\sen_2},\Set{\sen_3},\Set{\sen_4},\Set{\sen_4},\Set{\sen_6}}$ and thus $I_*(\Sen ; \Pub) = \log |\Pat_{\range{\Sen | \Pub}}^*| = \log |\Pat_{G_{\Pub}}^*| = \log 5$.
\end{example}

\section{Non-stochastic Privacy-Preserving Data Release}
\label{sec:Algo}

Instead of the original $\Pub$, the data curator publishes $\Rel$, a distorted or sanitized version of the public data $\Pub$. The problem is how to sanitize the data to preserve privacy, while keeping the utility of $\Rel$ above a certain level.
In this section, we formulate the non-stochastic privacy-preserving problem as the minimization of either of two privacy measures in Section~\ref{sec:PrivMeasure}, $L_0(\Sen \rightarrow \Rel)$ or $I_*(\Sen ; \Rel)$, subject to a utility constraint.
We consider the Lagrangian function of the non-stochastic privacy-preserving problem and propose two agglomerative clustering algorithms to determine  $\Rel$ for the minimization of $L_0(\Sen \rightarrow \Rel)$ and $I_*(\Sen ; \Rel)$, respectively.
We also consider the problem of finding a privacy-preserving data release scheme $\Rel$ that guarantees the non-stochastic indistinguishability, which is formulated as minimizing $L_0(\Sen \rightarrow \Rel)$ subject to $I_*(\Sen ; \Rel) = 0$ and the utility constraint, and show that this problem can also be solved by the proposed agglomerative clustering algorithm.

The main objective of the non-stochastic privacy-preserving problem is to determine $\range{\Rel}$ and the conditional range $\range{\Rel | \Pub}$ through a deterministic function $f \colon \range{\Pub} \mapsto \range{\Rel}$ with $|\range{\Pub}| \geq |\range{\Rel}|$.
Here, $f$ can be considered as a quantization function that clusters/merges all $\pub \in \range{\Pub | \rel}$ to $\rel = f(\pub)$.
We use a more convenient notation for $f$, the partition $\Qat = \Set{\range{\Pub | \rel} \colon \rel \in \range{\Rel}}$.
This privacy-preserving method is captured by the Markov chain $\Sen - \Pub - \Rel$, where, for a given joint range $\range{\Sen,\Pub}$, a sanitization method $\range{\Rel | \Pub}$ results in the conditional range
\begin{equation} \label{eq:QuantAux}
    \range{\Sen | \rel} = \bigcup_{\pub \in \range{\Pub | \rel}} \range{\Sen | \pub} = \range{\Sen | \PUB}, \quad \forall \rel \in \range{\Rel}
\end{equation}
for $\PUB = \range{\Pub | \rel} \in \Qat$.

\subsection{Minimizing Privacy Leakage $L_0(\Sen \rightarrow \Rel)$}
\label{subsec:MinL0}

For a given threshold $\theta$, consider the problem
\begin{equation} \label{eq:prob2}
    \min_{\range{\Rel|\Pub}} L_0(\Sen \rightarrow \Rel), \quad  \text{ s.t. } U_i(\Pub;\Rel) \geq \theta.
\end{equation}
That is, we aim to maximize the value of $k$ in $k$-anonymity (Lemma~\ref{lemma2}) subject to a constraint on the utility.
For the quantization $\Qat$ corresponding to $\range{\Rel | \Pub}$, we have $B_0(\Sen | \Rel) = \min_{\rel \in \range{\Rel}} \log |\range{\Sen | \rel}| = \min_{\PUB \in \Qat} \log |\range{\Sen | \PUB}|$. The Lagrangian function of \eqref{eq:prob2} is
$\Lat (\range{\Rel | \Pub},\lambda ) = - \min_{\PUB \in \Qat} \log |\range{\Sen | \PUB}| - \lambda U_i(\Pub ; \Rel)$ for all $\lambda \in [0,+\infty)$.

The utility of the released data $\Rel$ is measured by $U_i(\Pub ; \Rel)$. We consider two definitions:
$$ U_i(\Pub;\Rel) = \begin{cases} I_0(\Pub \rightarrow \Rel), & i = 1, \\ - \max_{\pub,\rel \colon \rel \in \range{\Rel | \pub}} d(\pub,\rel),  & i = 2.\end{cases}$$
The non-stochastic 0-information $I_0(\Pub \rightarrow \Rel) = H_0(\Pub) - \max_{\rel \in \range{\Rel}} \log |\range{\Pub | \rel}|$ denotes the minimal posterior entropy reduction on $\Pub$, which corresponds to coarsest resolution of $\Pub$ by observing $\Rel$. Therefore, $I_0(\Pub \rightarrow \Rel)$ indicates the worst-case data utility. So is $-\max_{\pub,\rel \colon \rel \in \range{\Rel | \pub}} d(\pub,\rel)$ for the pairwise distance/distortion\footnote{$d(\cdot,\cdot)$ could be the $\ell_p$-norm or any other pairwise dissimilarity measure. } function $d(\cdot,\cdot)$.

       \begin{algorithm} [tpb]
	       \label{algo:AgloCluster2}
	       \small
	       \SetAlgoLined
	       \SetKwInOut{Input}{input}\SetKwInOut{Output}{output}
	       \SetKwFor{For}{for}{do}{endfor}
            \SetKwRepeat{Repeat}{repeat}{until}
            \SetKwIF{If}{ElseIf}{Else}{if}{then}{else if}{else}{endif}
	       \BlankLine
           \Input{the Lagrangian multiplier $\lambda \in [0,+\infty)$ and the joint range $\range{\Sen,\Pub}$.}
	       \Output{the quantization $\Qat^{(t-1)}$.}
	       \BlankLine
                initiate $\Qat^{(0)} \coloneqq \Set{\Set{\pub} \colon \pub \in \range{\Pub}}$ and $t \coloneqq 0$\;
                \Repeat{ $\triangle  \Lat  \geq 0$ or $|\Qat^{(t)}| = 1$}{ \label{step:repeat1}
                    $ \Pi \coloneqq \argmin \Set{ \log |\range{\Sen | \PUB}| \colon \PUB \in \Qat^{(t)}} $ \;
                    $ \Qat^{(t + 1)} \coloneqq \Qat^{(t)}$\;
                    $t \coloneqq t + 1$\;
                    \Repeat{$\Pi = \emptyset $}{
                         $\PUB^* \coloneqq \argmax \big\{ U_i (\Pub ; \Rel_{\Set{\PUB,\PUB'}}^{(t)}) \colon \PUB' \in \Qat^{(t)}, \PUB' \neq \PUB, \range{\Sen|\PUB'} \neq \range{\Sen|\PUB}\big\}$ \label{step:Xopt2} \;
                        \nl$\Qat^{(t)} \coloneqq \Qat^{(t)}_{\Set{\PUB,\PUB^*}}$\;
                        $\Pi \coloneqq \Pi \setminus \Set{\PUB,\PUB^*}$\;
                    } \label{step:repeat2}
                    $\triangle \Lat  \coloneqq L(\range{\Rel^{(t)} | \Pub},\lambda ) - L(\range{\Rel^{(t-1)} | \Pub},\lambda )$\;
                }
                \Return $\Qat^{(t-1)}$\;
	   \caption{Agglomerative clustering algorithm for solving problem~\eqref{eq:prob2}}
	   \end{algorithm}

\subsubsection{Agglomerative Clustering Algorithm}
\label{subsec:Descent}

While problem~\eqref{eq:prob2} is a computationally complex optimization problem, in particular for large datasets, we propose a greedy agglomerative clustering in Algorithm~\ref{algo:AgloCluster2} that generates a locally optimal $\range{\Rel}$ and $\range{\Rel | \Pub}$ by iteratively merging the elements in $\range{\Pub}$.

Algorithm~\ref{algo:AgloCluster2} starts with $\Qat^{(0)} = \Set{\Set{x} \colon x \in \range{X}}$, which means no quantization.
In each iteration $t$, we strictly increase $\min_{\PUB \in \Qat^{(t)}} \log |\range{\Sen | \PUB}|$ by merging each $\PUB \in \argmin \Set{ \log |\range{\Sen | \PUB}| \colon \PUB \in \Qat^{(t)}}$ with another subset $\PUB^*$ that maximize the utility function $U_i (\Pub ; \Rel_{\Set{\PUB,\PUB'}}^{(t)})$ over the subsets $\PUB'$ in the current quantization $\Qat^{(t)}$.
We explain how to search $\PUB^*$ in step~\ref{step:Xopt2} as follows.

For each distinct pair $\PUB_1, \PUB_2 \in \Qat^{(t)}$, let the partition
$$ \Qat_{\Set{\PUB_1,\PUB_2}}^{(t)} = ( \Qat^{(t)} \setminus \Set{\PUB_1,\PUB_2} ) \cup \Set{\PUB_1 \sqcup \PUB_2}$$
be obtained by merging $\PUB_1$ and $\PUB_2$ and $\Rel_{\Set{\PUB_1,\PUB_2}}^{(t)}$ is the resulting uv.
Consider the utility function $U_1 (\Pub ;\Rel_{\Set{\PUB_1,\PUB_2}}^{(t)}) = I_0(\Pub \rightarrow \Rel_{\Set{\PUB_1,\PUB_2}}^{(t)}) = H_0(\Pub) - \max_{\PUB \in \Qat_{\Set{\PUB_1,\PUB_2}}^{(t)}} \log |\PUB|$.
Because merging $\PUB_1$ and $\PUB_2$ such that $\range{\Sen|\PUB_1} = \range{\Sen|\PUB_2} $ does not change the value of $L_0(\Sen \rightarrow \Rel^{(t)})$, we let
$$ \Phi(\Qat^{(t)}) = \Set{(\PUB_1,\PUB_2) \in \Qat^{(t)} \colon \PUB_1 \neq \PUB_2, \range{\Sen|\PUB_1} \neq \range{\Sen|\PUB_2}} $$
and consider
\begin{align*}
        (\PUB_1^*, \PUB_2^*) & \in  \argmax_{(\PUB_1,\PUB_2) \in \Phi(\Qat^{(t)})} U_1 (\Pub ;\Rel_{\Set{\PUB_1,\PUB_2}}^{(t)}) \\
        &= \argmin_{(\PUB_1,\PUB_2) \in \Phi(\Qat^{(t)})}   \ \max_{\PUB \in \Qat_{\Set{\PUB_1,\PUB_2}}^{(t)}} \log |\PUB|
\end{align*}
Here, we have $\max_{\PUB \in \Qat_{\Set{\PUB_1,\PUB_2}}^{(t)}} \log |\PUB| = \log \big( \max \Set{|\PUB_1| $ $+ |\PUB_2|, \max_{\PUB \in \Qat^{(t)}} |\PUB|} \big)$,
for which, if $|\PUB_1| + |\PUB_2| \leq $ $\max_{\PUB \in \Qat^{(t)} } |\PUB|, \forall (\PUB_1,\PUB_2) \in \Phi(\Qat^{(t)})$, then
$$\argmin_{(\PUB_1,\PUB_2) \in \Phi(\Qat^{(t)})} \max_{\PUB \in \Qat_{\Set{\PUB_1,\PUB_2}}^{(t)}} \log |\PUB| = \Phi(\Qat^{(t)}); $$
if there exists $ (\PUB_1,\PUB_2) \in \Phi(\Qat^{(t)})$ such that $|\PUB_1| + |\PUB_2| > \max_{\PUB \in \Qat^{(t)} } |\PUB|$, then
\begin{multline*}
    \argmin_{(\PUB_1,\PUB_2) \in \Phi(\Qat^{(t)})} \max_{\PUB \in \Qat_{\Set{\PUB_1,\PUB_2}}^{(t)}} \log |\PUB| \\ = \argmin_{(\PUB_1,\PUB_2) \in \Phi(\Qat^{(t)})} \Set{|\PUB_1| + |\PUB_2|}.
\end{multline*}
Therefore, we have $\argmin_{(\PUB_1,\PUB_2) \in \Phi(\Qat^{(t)})} \Set{|\PUB_1| + |\PUB_2|} \subseteq \argmin_{(\PUB_1,\PUB_2) \in \Phi(\Qat^{(t)})}   \max_{\PUB \in \Qat_{\Set{\PUB_1,\PUB_2}}^{(t)}} \log |\PUB| $. So, we can search for
$$
        (\PUB_1^*, \PUB_2^*)  \in \argmin_{(\PUB_1,\PUB_2) \in \Phi(\Qat^{(t)})} \big\{ |\PUB_1| + |\PUB_2| \big\}.
$$
Therefore, for the utility function $U_1 (\Pub ;\Rel_{\Set{\PUB,\PUB'}}^{(t)}) = I_0(\Pub \rightarrow \Rel_{\Set{\PUB,\PUB'}}^{(t)})$, we choose $\PUB^* \in \argmin \big\{ |\PUB \sqcup \PUB'| \colon \PUB' \in \Qat^{(t)}, \PUB' \neq \PUB, \range{\Sen|\PUB'} \neq \range{\Sen|\PUB}\big\}$.

For the utility function $U_2 = - \max_{\pub,\rel \colon \rel \in \range{\Rel | \pub}} d(\pub,\rel)$, a unique codeword $\rel$ is assigned to each cluster $\PUB \in \Qat$ such that $\PUB = \range{\Pub | \rel}$. Two possible choices are the centroid of the cluster $\rel = \frac{1}{|\PUB|}\sum_{\pub \in \PUB} \pub$ and $\rel = \pub$ for some $\pub \in \PUB$, but other choices are possible based on the application. We denote the maximum distortion for the cluster $\PUB \in \Qat$ by
   $$ \bar{d}(\PUB) = \max_{\pub \in \PUB} d(\pub,\rel)$$
and have
\begin{equation}
    \begin{aligned}
        (\PUB_1^*, \PUB_2^*) & \in \argmin_{(\PUB_1,\PUB_2) \in \Phi(\Qat^{(t)})}  \ \max_{\pub,\rel \colon \rel \in \range{\Rel^{(t)}_{\Set{\PUB_1,\PUB_2}} | \pub}} d(\pub,\rel) \\
        & = \argmin_{(\PUB_1,\PUB_2) \in \Phi(\Qat^{(t)})} \ \max_{\PUB \in \Qat^{(t)}_{\Set{\PUB_1,\PUB_2}}} \bar{d}(\PUB)
    \end{aligned} \nonumber
\end{equation}
Here, since $\max_{\PUB \in \Qat^{(t)}_{\Set{\PUB_1,\PUB_2}}} \bar{d}(\PUB) = \max \Set{\bar{d}(\PUB_1 \sqcup \PUB_2), \max_{\PUB \in \Qat^{(t)}} \bar{d}(\PUB)}$,
$\argmin_{(\PUB_1,\PUB_2) \in \Phi(\Qat^{(t)})} \bar{d}(\PUB_1 \sqcup \PUB_2) \subseteq \argmin_{(\PUB_1,\PUB_2) \in \Phi(\Qat^{(t)})} \max_{\PUB \in \Qat^{(t)}_{\Set{\PUB_1,\PUB_2}}} \bar{d}(\PUB)  $ and therefore we choose $\PUB^* \in \argmin \big\{ \bar{d}(\PUB \sqcup \PUB') \colon \PUB' \in \Qat^{(t)}, \PUB' \neq \PUB, \range{\Sen|\PUB'} \neq \range{\Sen|\PUB}\big\}$ in step~\ref{step:Xopt2}.

\begin{example} \label{ex:P2}
    For $\lambda = 0.3$, we run Algorithm~\ref{algo:AgloCluster2} to the uvs $\Sen$ and $\Pub$ in Example~\ref{ex:main} with the initiation $\Qat^{(0)} = \Set{\Set{\pub_1},\dotsc,\Set{\pub_7}}$. Consider the utility function $U_1(\Pub ; \Rel^{(t)}) = I_0(\Sen \rightarrow \Rel^{(t)})$.
    At the first iteration, we have $\Pi = \Set{\Set{\pub_2},\dotsc, \Set{\pub_7}}$, for which step~\ref{step:repeat1} to step~\ref{step:repeat2} produce the quantization $\Qat^{(1)} = \Set{\Set{\pub_1,\pub_2,\pub_7},\Set{\pub_3,\pub_5},\Set{\pub_4,\pub_6}}$ with the conditional ranges $\range{\Sen | \Set{\pub_1,\pub_2,\pub_7}} = \Set{\sen_1,\sen_2,\sen_6}$, $\range{\Sen | \Set{\pub_3,\pub_5}} = \Set{\sen_3,\sen_4}$ and $\range{\Sen | \Set{\pub_4,\pub_6}} = \Set{\sen_3,\sen_6}$. Since $\triangle \Lat = -\log 2 + \lambda \log 3 = -0.5245 < 0$, we continue the iterations.
    At the second iteration, we have $\Pi = \Set{\Set{\pub_{3},\pub_{5}},\Set{\pub_4,\pub_6}}$ and $\Qat^{(2)} = \Set{\Set{1,2,7},\Set{3,\dotsc,6}}$ after step~\ref{step:repeat2} with the conditional ranges $\range{\Sen | \Set{\pub_1,\pub_2,\pub_7}} = \Set{\sen_1,\sen_2,\sen_6}$ and $\range{\Sen | \Set{\pub_3,\dotsc,\pub_6}} = \Set{\sen_3,\sen_4,\sen_6}$, for which $\triangle \Lat = -0.4605 < 0$. We run the third iteration and have $\Qat^{(3)} = \Set{\Set{\pub_1,\dotsc,\pub_7}}$ returned.

    Let the values of $\pub \in \range{\Pub}$ be
    \begin{equation} \label{eq:values}
        \begin{aligned}
            & \pub_1 = 0.2,\ \pub_2 = 0.1,\ \pub_3 = 0.4,\ \pub_4 = 0.3,\\
            & \pub_5 = 0.6,\ \pub_6 = 1.5,\ \pub_7 = 1.
        \end{aligned}
    \end{equation}
    and run Algorithm~\ref{algo:AgloCluster2} for the utility function $U_2(\Pub ; \Rel^{(t)}) = - \max_{\pub,\rel \colon \rel \in \range{\Rel^{(t)} | \pub}} d(\pub,\rel)$ and $\lambda = 2.5$. At the end of 1st iteration, we have $\Pi = \Set{\Set{\pub_2},\dotsc, \Set{\pub_7}}$ and $\Qat^{(1)} = \Set{\Set{\pub_1,\pub_2,\pub_4}, \Set{\pub_3,\pub_5},\Set{\pub_6,\pub_7}}$ with the conditional ranges $\range{\Sen | \Set{\pub_1,\pub_2,\pub_4}} = \Set{\sen_1,\sen_2,\sen_3}$, $\range{\Sen | \Set{\pub_3,\pub_5}} = \Set{\sen_3,\sen_4}$ and $\range{\Sen | \Set{\pub_6,,\pub_7}} = \Set{\sen_5,\sen_6}$. Since $\triangle \Lat = - \log 2 + \lambda \bar{d}(\Set{\pub_6,\pub_7}) = -0.3750 < 0 $, we continue to iteration $t = 2$ where $\Pi = \Set{\Set{\pub_3,\pub_5},\Set{\pub_6, \pub_7}}$.
    But, at step~\ref{step:repeat2}, $\Qat^{(2)} = \Set{\Set{\pub_1,\pub_2,\pub_7},\Set{\pub_3,\dotsc,\pub_6}}$ but $\triangle \Lat = - \log 3 + \lambda (\bar{d}(\Set{\pub_1,\dotsc,\pub_7}) - \bar{d}(\Set{\pub_6,\pub_7})) = 0.0758 > 0$. Iteration terminates and the quantization $\Qat^{(1)} = \Set{\Set{\pub_1,\pub_2,\pub_7},\Set{\pub_3,\pub_5},\Set{\pub_4,\pub_6}}$ is returned.
    Algorithm~\ref{algo:AgloCluster2} ensures a strict reduction of the Lagrangian function in each iteration. See Fig.~\ref{fig:ConvergenceP2}.
\end{example}

\begin{figure}[tpb]
	\centering
    \scalebox{0.5}{\begin{tikzpicture}
\begin{axis}[
width=4.5in,
height=1.8in,
scale only axis,
xmin=0,
xmax=3,
xtick = {0,1,2,3},
xticklabels = {$0$,$1$,$2$,$3$},
xlabel={\large iteration index $t$},
ymin=0,
ymax=2.6,
ylabel={\large $\Lat(\range{\Rel^{(k)} | \Pub}, \lambda)$},
grid=major]
% \addplot[] is better than \addplot+[] here:
% it avoids scalings of the cycle list
\addplot [
line width = 1.5pt,
color=blue,
mark=*,
]
table[row sep=crcr]{
0 1.7428\\
1 1.282\\
2 0.7578\\
3 0\\
};
\addlegendentry{\Large $U_1$};

\addplot [
line width = 1.5pt,
color=red,
mark=square,
]
table[row sep=crcr]{
0 2.5850\\
1 2.21\\
};
\addlegendentry{\Large $U_2$};

\end{axis}
\end{tikzpicture} }
	\caption{The convergence performance in terms of $\Lat(\range{\Rel^{(t)} | \Pub},\lambda ) = - \min_{\PUB \in \Qat^{(t)}} \log |\range{\Sen | \PUB}| - \lambda U_i(\Pub ; \Rel^{(t)})$ versus the iteration index $t$ when Algorithm~\ref{algo:AgloCluster2} is applied to solve problem~\eqref{eq:prob2}. For the utility function $U_1(\Pub ; \Rel^{(t)}) = I_0(\Sen \rightarrow \Rel^{(t)})$ and $\lambda = 0.3$, the iteration terminates at $t = 3$; For the utility function $U_2(\Pub ; \Rel^{(t)}) = - \max_{\pub,\rel \colon \rel \in \range{\Rel^{(t)} | \Pub}} d(\pub,\rel)$ and $\lambda = 2.5$, the iteration terminates at $t = 1$. See Example~\ref{ex:P2}. Both plots are strictly decreasing.}
	\label{fig:ConvergenceP2}
\end{figure}

\subsection{Minimizing the Maximin Information $I_*$}

Consider the problem
\begin{equation} \label{eq:prob1}
    \min_{\range{\Rel|\Pub}} I_*(\Sen ; \Rel), \quad  \text{ s.t. } U_i(\Pub;\Rel) \geq \theta.
\end{equation}
The Lagrangian function of \eqref{eq:prob1} is
$ \Lat (\range{\Rel | \Pub},\lambda) = \log | \Pat_{G_{\Rel}}^* | - \lambda U_i(\Pub ; \Rel) $
for $\lambda \in [0, +\infty)$.
Here, $\Pat_{G_{\Rel}}^*$ is the finest decomposition of the undirected uncapacitated graph $G_{\Rel} = (\range{\Rel}, \Edge_{\range{\Sen | \Rel}})$ for the conditional range $\range{\Sen | \Rel}$.
Based on \eqref{eq:QuantAux} and Definition~\ref{def:OverPart}, we show below how to obtain $\Pat_{G_{\Rel}}$ from $\Pat_{G_{\Pub}}$ by a subgraph merging method, the idea of which is then used to propose the agglomerative clustering algorithm for solving problem~\eqref{eq:prob1}.

\subsubsection{Determining $\Pat_{G_{\Rel}}^*$}
\label{subsec:QtoG}

Initiate $\Pat_{G_{\Rel}}^* \coloneqq \Pat_{G_{\Pub}}^*$; For each $\PUB \in \Qat = \Set{\range{\Pub | \rel} \colon \rel \in \range{\Rel}}$, obtain all the subgraphs in $\Pat_{G_{\Rel}}^*$ that intersect with $\PUB$ as
$$ \GG_{\PUB} \coloneqq \Set{\G \in \Pat_{G_{\Rel}}^* \colon \G \cap \PUB \neq \emptyset} $$
and update $\Pat_{G_{\Rel}}^*$ by fusing all subgraphs in $\GG_{\PUB}$ to form one subset $\TGG_{\PUB} = \sqcup_{\G \in \GG_{\PUB}} \G$, i.e.,
$$ \Pat_{G_{\Rel}}^* \coloneqq (\Pat_{G_{\Rel}}^* \setminus \GG_{\PUB}) \cup \Set{\TGG_{\PUB}}. $$
The size $|\Pat_{G_{\Rel}}^*|$ is reducing in each iteration and $\Pat_{G_{\Rel}}^*$ is updated to the finest decomposition of $G_{\Rel} = (\range{\Rel}, \Edge_{\range{\Sen | \Rel}})$ at the end such that $I_*(\Sen ; \Rel) = \log | \Pat_{G_{\Rel}}^* |$.
For example, for $\Pat_{G_{\Pub}}^* = \Set{\Set{\pub_1,\pub_2},\Set{\pub_3,\pub_4},\Set{\pub_5},\Set{\pub_6},\Set{\pub_7}}$ in Fig.~\ref{fig:Gx} and the quantization $\Qat = \Set{\Set{\pub_1,\pub_3},\Set{\pub_2,\pub_5},\Set{\pub_4,\pub_6},\Set{\pub_7}}$, we get the finest decomposition $\Pat_{G_{\Rel}} = \Set{\Set{\pub_1,\dotsc,\pub_6},\Set{\pub_7}}$ at the end of the above iterative process (see Fig.~\ref{fig:II}(c)).

\subsubsection{Agglomerative Clustering Algorithm}

We propose an agglomerative clustering algorithm in Algorithm~\ref{algo:AgloCluster1} for problem \eqref{eq:prob1}, which can be interpreted as an iterative subgraph merging process.
Algorithm~\ref{algo:AgloCluster1} starts with $\Qat^{(0)} = \Set{\Set{x} \colon x \in \range{X}}$.
In each iteration $t$, we have the undirected graph $G$ updated to $G_{\Rel^{(t)}} = (\range{\Rel^{(t)}}, \Edge_{\range{\Sen | \Rel^{(t)}}})$ and the partition $\Pat^{(t)}$ updated to its finest decomposition $\Pat_{G_{\Rel^{(t)}}}^*$. Here, $\Rel^{(t)}$ denotes the released data corresponding to the quantization $\Qat^{(t)}$.
Therefore, $G$ and $\Pat^{(0)}$ are initiated as $G_{\Pub} = (\range{\Pub}, \Edge_{\range{\Sen | \Pub}})$ and $\Pat_{G_{\Pub}}^*$, respectively.

Algorithm~\ref{algo:AgloCluster1} iteratively merge two subsets $\PUB_1^*, \PUB_2^* \in \Qat^{(t)}$ that reduce the Lagrangian function $\Lat (\range{\Rel^{(t)} | \Pub},\lambda)$ the most (step~\ref{step:deltaL1} to step~\ref{step:deltaL2}).
This is done as follows. Consider the Lagrangian function $ \Lat (\range{\Rel^{(t)} | \Pub},\lambda) = \log | \Pat_{G_{\Rel^{(t)}}}^* | - \lambda U_i(\Pub ; \Rel^{(t)})$.
Since merging subsets of $\Qat^{(t)}$ that belong to the same subgraph in $\Pat^{(t)} = \Pat_{G_{\Rel^{(t)}}}^* $ does not reduce $| \Pat_{G_{\Rel^{(t)}}}^* |$ but only reduce utility $U_i(\Pub ; \Rel^{(t)})$, we can limit the merging operation to the subsets that belong to disconnected subgraphs in $\Qat^{(t)}$. To this end, the set of candidate subsets of $\Qat^{(t)}$ for the pairwise merging operation at iteration $t$ is defined as
\begin{multline}
    \Psi(\Qat^{(t)},\Pat^{(t)}) = \big\{ (\PUB_1,\PUB_2) \colon \PUB_1,\PUB_2 \in \Qat^{(t)}, \PUB_1 \neq \PUB_2, \\
                        \exists \G_1, \G_2 \in \Pat^{(t)}, \G_1 \cap \PUB_1 \neq \emptyset, \G_2 \cap \PUB_2 \neq \emptyset, \G_1 \neq \G_2 \big\}. \nonumber
\end{multline}

       \begin{algorithm} [t]
	       \label{algo:AgloCluster1}
	       \small
	       \SetAlgoLined
	       \SetKwInOut{Input}{input}\SetKwInOut{Output}{output}
	       \SetKwFor{For}{for}{do}{endfor}
            \SetKwRepeat{Repeat}{repeat}{until}
            \SetKwIF{If}{ElseIf}{Else}{if}{then}{else if}{else}{endif}
	       \BlankLine
           \Input{the Lagrangian multiplier $\lambda \in [0,+\infty)$ and the joint range $\range{\Sen,\Pub}$.}
	       \Output{the quantization $\Qat^{(t)}$, the partition $\Pat^{(t)}$ that determines the maximin information $I_*(\Sen; \Rel^{(t)}) = \log |\Pat^{(t)}|$ and the graph $G$ that equals $G_{\Rel^{(t)}} = (\range{\Rel^{(t)}}, \Edge_{\range{\Sen | \Rel^{(t)}}})$.}
	       \BlankLine
                initiate the quantization $\Qat^{(0)} \coloneqq \Set{\Set{\pub} \colon \pub \in \range{\Pub}}$, the graph $G \coloneqq G_{\Pub}$ and the finest decomposition $\Pat^{(0)} \coloneqq \Pat_{G_{\Pub}}^*$\;
                $t \coloneqq 0$\;
                \While{ $|\Pat^{(t)}| > 1$}{
                    Choose a maximizer $(\PUB_1^*, \PUB_2^*)$ of  \label{step:deltaL1}
                    \begin{equation} \label{eq:Xopt1}
                            \max_{(\PUB_1,\PUB_2) \in  \Psi(\Qat^{(t)}, \Pat^{(t)})}  U_i (\Pub ; \Rel_{\Set{\PUB_1,\PUB_2}}^{(t)}) ;
                    \end{equation}
                    \nl $\triangle \Lat \coloneqq \log \frac{|\Pat^{(t)}| - 1}{|\Pat^{(t)}|} - \lambda ( U_i (\Pub; \Rel^{(t)}_{\Set{\PUB_1^*,\PUB_2^*}}) - U_i (\Pub ; \Rel^{(t)}))$\;
                    \lIf{$\triangle \Lat  \geq 0$}{break} \label{step:deltaL2}
                    Merge nodes $\PUB_1^*$ and $\PUB_2^*$ in graph $G$\;
                    $\Qat^{(t+1)} \coloneqq \Qat^{(t)}_{\Set{\PUB_1^*,\PUB_2^*}}$\;
                    $\Pat^{(t+1)} \coloneqq \Pat^{(t)}_{\Set{\G_1^*,\G_2^*}}$, where $\G_1^*, \G_2^* \in \Pat^{(t)}$ such that $\G_1^* \cap \PUB_1^* \neq \emptyset$ and $\G_2^* \cap \PUB_2^* \neq \emptyset$\;
                    $t \coloneqq t + 1$\;
                }
                \Return $\Qat^{(t)}$, $\Pat^{(t)}$ and $G$\;
	   \caption{Agglomerative clustering algorithm for solving problem~\eqref{eq:prob1}}
	   \end{algorithm}

We determine the steepest decent direction $(\PUB_1^*, \PUB_2^*)$ as the maximizer of $\max U_i (\Pub ; \Rel_{\Set{\PUB_1,\PUB_2}}^{(t)})$ over $\Psi(\Qat^{(t)},\Pat^{(t)})$.
As explained Section~\ref{subsec:Descent}, for the utility function $U_1 (\Pub ;\Rel_{\Set{\PUB_1,\PUB_2}}^{(t)}) = I_0(\Pub \rightarrow \Rel_{\Set{\PUB_1,\PUB_2}}^{(t)})$, we choose
\begin{equation} \label{eq:prob1U1}
    \begin{aligned}
        (\PUB_1^*, \PUB_2^*) &
                             & \in \argmin_{(\PUB_1,\PUB_2) \in \Psi(\Qat^{(t)},\Pat^{(t)})} \big\{ |\PUB_1| + |\PUB_2| \big\}.
    \end{aligned}
\end{equation}
If the minimizer of \eqref{eq:prob1U1} is not unique, we should choose the pair $(\PUB_1^*, \PUB_2^*)$ that connects the largest two subgraphs in $\Pat_{G_{\Rel^{(t)}}}^*$. The purpose is to have a smaller minimum of \eqref{eq:prob1U1} in the subsequent iterations $t' > t$. See the example below.
For the utility function $U_2 = - \max_{\pub,\rel \colon \rel \in \range{\Rel | \pub}} d(\pub,\rel)$, we chose
$$(\PUB_1^*, \PUB_2^*) \in \argmin_{(\PUB_1,\PUB_2) \in \Psi(\Qat^{(t)},\Pat^{(t)})} \bar{d}(\PUB_1 \sqcup \PUB_2).$$

\begin{figure}[t]
	\centering
        \subfigure[$t = 1$]{\scalebox{0.65}{\begin{tikzpicture}

%nodes
\draw (0.5,1) circle (0.3);
\node at (0.5,1) {\Large $\pub_2$};

\draw (0.5,-1)  circle (0.3);
\node at (0.5,-1) {\Large $\pub_4$};

\draw (-1.5,1) [color = blue] ellipse (0.4 and 0.3);
\node at (-1.5,1) [color = blue] {\Large $\pub_{13}$};

% \draw (-1.5,-1) circle (0.3);
% \node at (-1.5,-1) {\Large $3$};

\draw (2,0.7) circle (0.3);
\node at (2,0.7) {\Large $\pub_5$};

\draw (2,-0.7) circle (0.3);
\node at (2,-0.7) {\Large $\pub_6$};

\draw (3.5,0) circle (0.3);
\node at (3.5,0) {\Large $\pub_7$};

% edges
\draw (-1.1,1) --  (0.2,1);
\draw (0.29,-0.8) -- (-1.2,0.8);

\end{tikzpicture} }} \qquad\quad
        \subfigure[$t = 2$]{\scalebox{0.65}{\begin{tikzpicture}

%nodes
\draw (0.5,1) [color = blue] ellipse (0.4 and 0.3);
\node at (0.5,1) [color = blue] {\Large $\pub_{25}$};

\draw (0.5,-1) circle (0.3);
\node at (0.5,-1) {\Large $\pub_4$};

\draw (-1.5,1) ellipse (0.4 and 0.3);
\node at (-1.5,1) {\Large $\pub_{13}$};

\draw (2,-0.7) circle (0.3);
\node at (2,-0.7) {\Large $\pub_6$};

\draw (3,0) circle (0.3);
\node at (3,0) {\Large $\pub_7$};

% edges
\draw (-1.1,1) --  (0.1,1);
\draw (0.29,-0.8) -- (-1.2,0.8);

\end{tikzpicture} }} \\
        \subfigure[$t = 3$]{\scalebox{0.65}{\begin{tikzpicture}

%nodes
\draw (0.5,1) ellipse (0.4 and 0.3);
\node at (0.5,1) {\Large $\pub_{25}$};

\draw (0.5,-1) [color = blue] ellipse (0.4 and 0.3);
\node at (0.5,-1) [color = blue] {\Large $\pub_{46}$};

\draw (-1.5,1) ellipse (0.4 and 0.3);
\node at (-1.5,1) {\Large $\pub_{13}$};

\draw (2.5,0) circle (0.3);
\node at (2.5,0) {\Large $\pub_7$};

% edges
\draw (-1.1,1) --  (0.1,1);
\draw (0.29,-0.738) -- (-1.2,0.75);

\end{tikzpicture} }} \qquad\quad
        \subfigure[$t = 4$]{\scalebox{0.65}{\begin{tikzpicture}

%nodes
\draw (0.5,1) ellipse (0.4 and 0.3);
\node at (0.5,1) {\Large $\pub_{25}$};

\draw (0.5,-1) ellipse (0.4 and 0.3);
\node at (0.5,-1) {\Large $\pub_{46}$};

\draw (-1.5,1) [color = blue] ellipse (0.5 and 0.3);
\node at (-1.5,1) [color = blue] {\Large $\pub_{137}$};

% edges
\draw (-1,1) --  (0.1,1);
\draw (0.29,-0.738) -- (-1.2,0.75);

\end{tikzpicture} }}
	\caption{In Example~\ref{ex:P1U1}, the change of the undirected uncapacitated graph $G$ at the end of each iteration $t$ of Algorithm~\ref{algo:AgloCluster1}, where $|\Pat^{(t)}|$, the number of disconnected subgraphs in $G$, is reduced from $t =5$ to $t = 1$ and so is $I_*(\Sen ; \Rel^{(t)})$ from $\log_2 5$ to $0$. The subscription of a node in $G$ denotes the original elements in $\range{\Pub}$ that are fused to form this super-node, e.g., $\pub_{13}$ is the fusion of $\pub_1$ and $\pub_3$.  }
	\label{fig:II}
\end{figure}
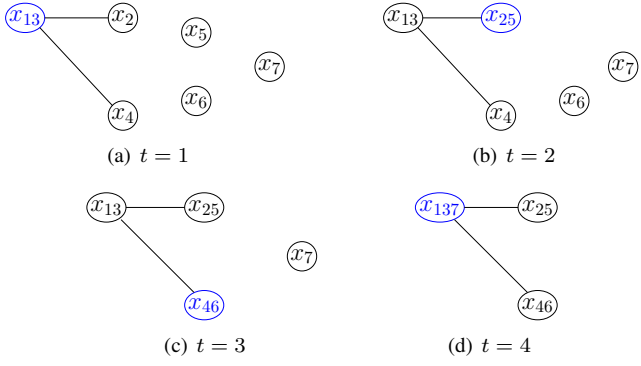

\begin{example} \label{ex:P1U1}
    For $\lambda = 0.3$,
    we apply Algorithm~\ref{algo:AgloCluster1} to the uvs $\Sen$ and $\Pub$ with the joint range $\range{\Sen, \Pub}$ in Example~\ref{ex:main} to solve the problem~\eqref{eq:prob1} for $U_1(\Pub ; \Rel) = I_0(\Sen \rightarrow \Rel)$.
 We initiate $\Qat^{(0)} = \Set{\Set{\pub_1},\dotsc,
    \Set{\pub_7}}$, $G$ as the undirected graph in Fig.~\ref{fig:Gx} and $\Pat^{(0)} = \Set{\Set{\pub_1,\pub_2},\Set{\pub_3,\pub_4},\Set{\pub_5},\Set{\pub_6},\Set{\pub_7}}$. Since $(\Set{\pub_1},\Set{\pub_2})$ and $(\Set{\pub_3},\Set{\pub_4})$ are already connected, candidates for pairwise merging are
    \begin{align*}
        \Psi(\Qat^{(0)},\Pat^{(0)}) = &\big\{ (\Set{\pub},\Set{\pub'}) \colon \pub,\pub' \in \range{\Pub},  \pub \neq \pub' \big\}\\ &\setminus  \big\{ (\Set{\pub_1},\Set{\pub_2}), (\Set{\pub_3},\Set{\pub_4}) \big\}. \nonumber
    \end{align*}
Note that $|\PUB_1| + |\PUB_2| = 2$ for all $(\PUB_1,\PUB_2) \in \Psi(\Qat^{(0)},\Pat^{(0)}) $.
    We choose to merge $(\Set{\pub_1},\Set{\pub_3})$ that connect the large two subgraphs $\Set{\pub_1,\pub_2}$ and $\Set{\pub_3,\pub_4}$ in the undirected graph in Fig.~\ref{fig:Gx}. This results in a strict reduction of the Lagrangian function $\triangle \Lat = \log_2 \frac{4}{5} - \lambda \log_2 \frac{2}{1} = -0.0219$.\footnote{One can show that, if $\Set{\pub_5}$ $\Set{\pub_6}$ is merged at the 1st iteration, we have $\min_{(\PUB_1,\PUB_2) \in \Psi(\Qat^{(2)},\Pat^{(2)})} \{ |\PUB_1| + |\PUB_2| \} = 3$ at the 3rd iteration. But in this example, the minimum is $2$ when $t = 3$. }
    We get the new quantization $\Qat^{(1)} = \Set{\Set{\pub_1,\pub_3},\Set{\pub_2},\Set{\pub_4},\dotsc,\Set{\pub_7}}$ and the updated graph $G$ in Fig.~\ref{fig:II}(a) with the finest decomposition being $\Pat^{(1)} = \Set{\Set{\pub_1,\dotsc,\pub_4},\Set{\pub_5},\Set{\pub_6},\Set{\pub_7}}$.
    At the 2nd iteration,
    \begin{multline}
        \Psi(\Qat^{(1)},\Pat^{(1)}) = \big\{ (\PUB_1,\PUB_2) \in \Qat^{(1)} \colon  \PUB_1 \neq \PUB_2 \big\} \setminus \\ \big\{ (\Set{\pub_{1},\pub_{3}},\Set{\pub_2}), (\Set{\pub_{1},\pub_{3}},\Set{\pub_4}),(\Set{\pub_2},\Set{\pub_4}) \big\}. \nonumber
    \end{multline}
    Among the minimizers in $\argmin \big\{ |\PUB_1| + |\PUB_2| \colon (\PUB_1,\PUB_2) \in \Psi(\Qat^{(1)},\Pat^{(1)}) \big\}$, we choose $(\Set{2},\Set{5})$ that connect the largest two subgraphs $\Set{1,\dotsc,4}$ and $\Set{5}$ and result in $\triangle \Lat < 0$. The graph $G$ is updated to Fig.~\ref{fig:II}(b). We have $\Qat^{(2)} = \Set{\Set{\pub_1,\pub_3},\Set{\pub_2,\pub_5},\Set{\pub_4},\Set{\pub_6},\Set{\pub_7}}$ and $\Pat^{(2)} = \Set{\Set{\pub_1,\dotsc,\pub_5},\Set{\pub_6},\Set{\pub_7}}$.
    Repeating in the same way, we have $\Qat^{(3)} = \Set{\Set{\pub_1,\pub_3},\Set{\pub_2,\pub_5},\Set{\pub_4,\pub_6},\Set{\pub_7}}$, $\Pat^{(3)} = \Set{\Set{\pub_1,\dotsc,\pub_6},\Set{\pub_7}}$ and the updated graph $G$ in Fig.~\ref{fig:II}(c) at the end of the 3rd iteration.
    At the 4th iteration, $\Qat^{(4)} = \Set{\Set{\pub_1,\pub_3,\pub_7},\Set{\pub_2,\pub_5},\Set{\pub_4,\pub_6}}$, $\Pat^{(4)} = \Set{\Set{\pub_1,\dotsc,\pub_7}}$ are returned.
\end{example}

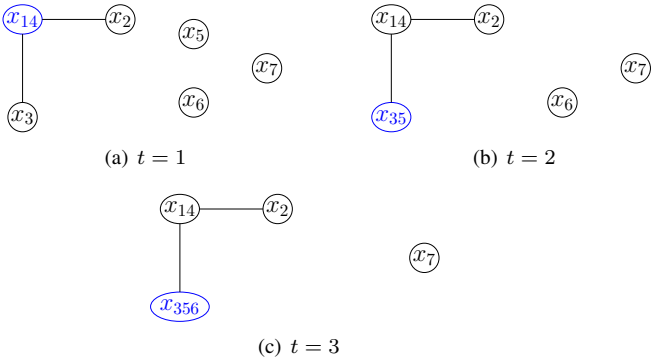
\begin{figure}[t]
	\centering
        \subfigure[$t = 1$]{\scalebox{0.65}{\begin{tikzpicture}

%nodes
\draw (0.5,1) circle (0.3);
\node at (0.5,1) {\Large $\pub_2$};

\draw (-1.5,1) [color = blue] ellipse (0.4 and 0.3);
\node at (-1.5,1) [color = blue] {\Large $\pub_{14}$};

\draw (-1.5,-1) circle (0.3);
\node at (-1.5,-1) {\Large $\pub_3$};

\draw (2,0.7) circle (0.3);
\node at (2,0.7) {\Large $\pub_5$};

\draw (2,-0.7) circle (0.3);
\node at (2,-0.7) {\Large $\pub_6$};

\draw (3.5,0) circle (0.3);
\node at (3.5,0) {\Large $\pub_7$};

% edges
\draw (-1.1,1) --  (0.2,1);
\draw (-1.5,0.7) -- (-1.5,-0.7);

\end{tikzpicture} }}\qquad\quad
        \subfigure[$t = 2$]{\scalebox{0.65}{\begin{tikzpicture}

%nodes
\draw (0.5,1) circle (0.3);
\node at (0.5,1) {\Large $\pub_2$};

\draw (-1.5,1) ellipse (0.4 and 0.3);
\node at (-1.5,1) {\Large $\pub_{14}$};

\draw (-1.5,-1) [color = blue] ellipse (0.4 and 0.3);
\node at (-1.5,-1) [color = blue] {\Large $\pub_{35}$};

\draw (2,-0.7) circle (0.3);
\node at (2,-0.7) {\Large $\pub_6$};

\draw (3.5,0) circle (0.3);
\node at (3.5,0) {\Large $\pub_7$};

% edges
\draw (-1.1,1) --  (0.2,1);
\draw (-1.5,0.7) -- (-1.5,-0.7);

\end{tikzpicture} }} \\
        \subfigure[$t = 3$]{\scalebox{0.65}{\begin{tikzpicture}

%nodes
\draw (0.5,1) circle (0.3);
\node at (0.5,1) {\Large $\pub_2$};

\draw (-1.5,1) ellipse (0.4 and 0.3);
\node at (-1.5,1) {\Large $\pub_{14}$};

\draw (-1.5,-1) [color = blue] ellipse (0.6 and 0.3);
\node at (-1.5,-1) [color = blue] {\Large $\pub_{356}$};

\draw (3.5,0) circle (0.3);
\node at (3.5,0) {\Large $\pub_7$};

% edges
\draw (-1.1,1) --  (0.2,1);
\draw (-1.5,0.7) -- (-1.5,-0.7);

\end{tikzpicture} }} \qquad\qquad
	\caption{In Example~\ref{ex:P1U2}, the change of the undirected uncapacitated graph $G$ of the problem at the end of each iteration $t$ of Algorithm~\ref{algo:AgloCluster1}, where $|\Pat^{(t)}|$, the number of connected subgraphs is reduced from $|\Pat^{(0)}| = 5$ (see Fig. \ref{fig:Gx}) to $|\Pat^{(3)}| = 2$.}
	\label{fig:ID}
\end{figure}

\begin{example} \label{ex:P1U2}
For the values of $\Pub$ in \eqref{eq:values}, we reset $\pub_7 = 4$. Consider the utility function $U_2 = - \max_{\pub,\rel \colon \rel \in \range{\Rel | \pub}} d(\pub,\rel)$.
We choose the centroid $\rel = \frac{1}{|\PUB|}\sum_{\pub \in \PUB} \pub$ for $\PUB = \range{\Pub | \rel}$ and $d(\pub , \rel) = \| \pub - \rel \|_2 = |\pub - \rel|$.
For $\lambda = 0.3$, we apply Algorithm~\ref{algo:AgloCluster1} to the uvs $\Sen$ and $\Pub$ with the joint range $\range{\Sen, \Pub}$ in Example~\ref{ex:main} to solve the problem~\eqref{eq:prob1}.
We have the graph $G$ updated as in Fig.~\ref{fig:ID}. At the end of the third iteration, $\Qat^{(3)} = \Set{\Set{\pub_1,\pub_4},\Set{\pub_3,\pub_5,\pub_6},\Set{\pub_2},\Set{\pub_7}}$, $\Pat^{(3)} = \Set{\Set{\pub_1,\dotsc,\pub_6},\Set{\pub_7}}$ are returned. In Fig.~\ref{fig:ConvergenceP1}, we plot the convergence performance of Algorithm~\ref{algo:AgloCluster1} in Examples~\ref{ex:P1U1} and \ref{ex:P1U2}. It can be seen that Algorithm~\ref{algo:AgloCluster1} ensures a strict reduction of the Lagrangian function $L(\range{\Rel^{(t)} | \Pub},\lambda)$ in each iteration $t$.
\end{example}

\begin{figure}[tpb]
	\centering
    \scalebox{0.5}{\begin{tikzpicture}
\begin{axis}[
width=4.5in,
height=1.8in,
scale only axis,
xmin=0,
xmax=4,
xtick = {0,1,2,3,4},
xticklabels = {$0$,$1$,$2$,$3$,$4$},
xlabel={\large iteration index $t$},
ymin=0,
ymax=3.5,
ylabel={\large $\Lat(\range{\Rel^{(k)} | \Pub}, \lambda)$},
grid=major]
% \addplot[] is better than \addplot+[] here:
% it avoids scalings of the cycle list
\addplot [
line width = 1.5pt,
color=blue,
mark=*,
]
table[row sep=crcr]{
0 3.1641\\
1 2.5422\\
2 2.1272\\
3 1.5422\\
4 0.3667\\
};
\addlegendentry{\Large $U_1$};

\addplot [
line width = 1.5pt,
color=red,
mark=square,
]
table[row sep=crcr]{
0 2.3219\\
1 2.0015\\
2 1.6150\\
3 1.2001\\
};
\addlegendentry{\Large $U_2$};

\end{axis}
\end{tikzpicture} }
	\caption{The convergence performance in terms of $\Lat(\range{\Rel^{(t)} | \Pub},\lambda) = \log | \Pat^{(t)} | - \lambda U_i(\Pub ; \Rel^{(t)})$ versus the iteration index $t$ when Algorithm~\ref{algo:AgloCluster1} is applied to solve problem~\eqref{eq:prob1} for $\lambda = 0.3$. For the utility function $U_1(\Pub ; \Rel^{(t)}) = I_0(\Sen \rightarrow \Rel^{(t)})$, the iteration terminates at $t = 4$ (see Example~\ref{ex:P1U1}); For the utility function $U_2(\Pub ; \Rel^{(t)}) = -\max_{\pub,\rel \colon \rel \in \range{\Rel^{(t)} | \pub}} d(\pub,\rel)$, the iteration terminates at $t = 3$ (see Example~\ref{ex:P1U2}). As guaranteed by Algorithm~\ref{algo:AgloCluster1}, both plots are strictly decreasing. }
	\label{fig:ConvergenceP1}
\end{figure}

\subsection{Minimizing $L_0(\Sen \rightarrow \Rel)$ and Restricting $I_*(\Sen ; \Pub) = 0$}

As discussed in Section~\ref{subsec:PrivRelation}, we have $L_0(\Sen \rightarrow \Rel) - I_*(\Sen; \Rel) \geq 0$, where $I_*(\Sen; \Rel)$ can be considered as the most bits in the leakage $L_0(\Sen \rightarrow \Rel)$ that can be estimated/distinguished by an adversary without error.
Also, due to the fact that the privacy leakage $L_0(\Sen \rightarrow \Rel)$ could be nonzero when $I_*(\Sen; \Rel) = 0$, it is worth considering the problem of how to find a data release scheme that ensures the non-stochastic indistinguishability
\begin{equation} \label{eq:prob3}
    \min_{\range{\Rel|\Pub}} L_0(\Sen \rightarrow \Rel), \ \text{ s.t. } I_*(\Sen; \Pub) =0, U_i(\Pub;\Rel) \geq \theta.
\end{equation}
%\farhad{[This problem might always be feasible as the utility constraint might contradict the $I_*=0$. Should we comment on this? This translate to not being able to find $|\Pat^{(t)}| = 1$ in your discussion below.]}
In \eqref{eq:prob3}, we assume that $\theta \geq \min_{\range{\Rel | \Pub} \colon I_*(\Sen ; \Rel) = 0} U_i(\Pub;\Rel) $.\footnote{This is to ensure a nonempty feasible region for \eqref{eq:prob3}. The proposed clustering algorithm based on the Lagrangian function \eqref{eq:LagrangianProb3} does not require the value of $\min_{\range{\Rel | \Pub} \colon I_*(\Sen ; \Rel) = 0} U_i(\Pub;\Rel)$ in advance, since the returned solution $\Rel$ for any given $\lambda$ is always feasible.}
The Lagrangian function is
\begin{multline} \label{eq:LagrangianProb3}
    \Lat(\range{\Sen | \Pub},\lambda',\lambda) = \\ - \min_{\PUB \in \Qat} \log |\range{\Sen | \PUB}| + \lambda' \log |\Pat_{G_{\Rel}}| - \lambda U_i(\Pub ; \Rel)
\end{multline}
for $\lambda',\lambda \geq 0$.
The same as Algorithm~\ref{algo:AgloCluster1}, the steepest decent direction $(\PUB_1^*, \PUB_2^*)$ can still be searched over the set $\Psi(\Pat^{(t)},\Qat^{(t)})$, but the iteration should repeat until $|\Pat^{(t)}| = 1$.
The resulting algorithm is the same as Algorithm~\ref{algo:AgloCluster1} except that steps~\ref{step:deltaL1} to \ref{step:deltaL2} are replaced by
    \begin{equation} \label{eq:ObjProb3}
        (\PUB_1^*, \PUB_2^*) \coloneqq  \argmin_{(\PUB_1,\PUB_2) \in  \Psi(\Pat^{(t)},\Qat^{(t)}) } \triangle \Lat ,
    \end{equation}
    where
                    \begin{equation}
                        \begin{aligned}
                            \triangle \Lat  & =  -\min_{\PUB \in \Qat^{(t)}_{\Set{\PUB_1,\PUB_2}}} \log |\range{\Sen | \PUB}| + \min_{\PUB \in \Qat^{(t)}} \log |\range{\Sen | \PUB}|  \\
                                         & \qquad +  \lambda (U_i(\Pub ; \Rel^{(t)}) - U_i(\Pub ; \Rel^{(t)}_{\Set{\PUB_1,\PUB_2}})).
                        \end{aligned}\nonumber
                    \end{equation}
If the minimizers of \eqref{eq:ObjProb3} is not unique, we choose $(\PUB_1^*, \PUB_2^*)$ with the minimum $|\range{\Sen | \PUB_1^*}|$ and $|\range{\Sen | \PUB_2^*}|$.
Note, the above method ensures a strict reduction of $I_*(\Sen; \Rel^{(t)})$ for each iteration $t$, which is equivalent to the case $\lambda' \gg \lambda$.

\begin{example} \label{ex:P3}
    We apply the above method to the uvs $\Sen$ and $\Pub$ in Example~\ref{ex:main} for the utility function $U_1(\Pub ; \Rel) = I_0(\Sen \rightarrow \Rel$ and $\lambda = 0.3$. Starting with $\Qat^{(0)} = \Set{\Set{\pub_1},\dotsc,
\Set{\pub_7}}$ and $\Pat^{(0)} = \Set{\Set{\pub_1,\pub_2},\Set{\pub_3,\pub_4},\Set{\pub_5},\Set{\pub_6},\Set{\pub_7}}$, at the first iteration, we have $\argmin \Set{ \triangle L \colon (\PUB_1,\PUB_2) \in  \Psi(\Pat^{(0)},\Qat^{(0)}) } = \Psi(\Pat^{(0)},\Qat^{(0)})$, where we choose $(\PUB_1^*,\PUB_2^*) = (\Set{2},\Set{3})$ such that $|\range{\Sen | \Set{2}}| = |\range{\Sen | \Set{3}}| = 2$ are the largest subgraphs.
We then have $\Qat^{(1)} = \Set{\Set{\pub_1},\Set{\pub_2,\pub_3},\Set{\pub_4},\dotsc,\Set{\pub_7}}$ and $\Pat^{(1)} = \Set{\Set{\pub_1,\dotsc,\pub_4},\Set{\pub_5},\Set{\pub_6},\Set{\pub_7}}$.
At the second iteration, we choose $(\PUB_1^*,\PUB_2^*) = (\Set{\pub_4},\Set{\pub_5})$ and have $\Qat^{(2)} = \Set{\Set{\pub_1},\Set{\pub_2,\pub_3},\Set{\pub_4,\pub_5},\Set{\pub_6},\Set{\pub_7}}$ and $\Pat^{(2)} = \Set{\Set{\pub_1,\dotsc,\pub_5},\Set{\pub_6},\Set{\pub_7}}$.
At the third iteration, we have $(\PUB_1^*,\PUB_2^*) = (\Set{\pub_6},\Set{\pub_7}) $ so that $\Qat^{(3)} = \Set{\Set{\pub_1},\Set{\pub_2,\pub_3},\Set{\pub_4,\pub_5},\Set{\pub_6,\pub_7}}$ and $\Pat^{(3)} = \Set{\Set{\pub_1,\dotsc,\pub_5},\Set{\pub_6,\pub_7}}$.
At the fourth iteration, there are only two subgraphs $\Set{\pub_1,\dotsc,\pub_5}$ and $\Set{\pub_6,\pub_7}$ in $G_{\Rel^{(t)}}$. We have $(\PUB_1^*,\PUB_2^*) = (\Set{\pub_1},\Set{\pub_6,\pub_7})$ so that $|\Pat^{(4)}| = |\Set{\Set{\pub_1,\dotsc,\pub_7}}| = 1$ finally. The quantization method $\Qat^{(4)} = \Set{\Set{\pub_1,\pub_6,\pub_7},\Set{\pub_2,\pub_3},\Set{\pub_4,\pub_5}}$ is returned.

\end{example}

Note, when the sensitive data $\Sen$ is also observable or available, the data curator can sanitize the data by quantizing the joint range $\range{\Sen,\Pub}$. It can be modeled by Markov chain $\Sen - Y - \hat{Y}$, where $Y = (\Sen,\Pub)$ denotes the public data and $\hat{Y}$ is its sanitization. The purpose is to search a privacy-preserving quantization function $f \colon \range{Y} \mapsto \range{\hat{Y}}$ that determines $\range{\hat{Y} | Y}$, for which, the proposed algorithms in this section also apply by replacing $\Pub$ and $\Rel$ with $Y$ and $\hat{Y}$, respectively.

It is also easy to extend to a system model where the adversary has side information $A$.
With the range of the sensitive data $\Sen$ conditioned on $A$ before and after the data release $\hat{X}$ denoted by $\range{\Sen | A}$ and $\range{\Sen | \Rel, A}$, respectively, the privacy leakage $L_0(\Sen \rightarrow \Rel|A) = \max_{(\rel,a) \in \range{\Rel,A}} \log \frac{|\range{\Sen | a}|}{|\range{\Sen | \rel,a}|}$ and the maximin information $I_*(\Sen;\Rel,A) = \log |\Pat_{\range{\Sen|\Rel,A}}^*|$ can be used as the privacy measures. Here, $\Pat_{\range{\Sen|\Rel,A}}^*$ refers to the finest decomposition of the undirected graph $G_{\Pub,A} = (\range{\Pub,A}, \Edge_{\range{\Sen | \Pub,A}})$.
In this case, we are instead considering the problem $\min_{\range{\Rel|\Pub}} L_0(\Sen \rightarrow \Rel|A) \text{ or }I_*(\Sen ; \Rel, A), \text{ s.t. } U_i(\Pub;\Rel) \geq \theta$, the solution to which can also be searched by Algorithm~\ref{algo:AgloCluster2} or \ref{algo:AgloCluster1}.

\section{Notes on Stochastic Information-theoretic Privacy and Differential Privacy}
\label{sec:ToITPriv}

The studies on privacy in traditional (i.e., stochastic) information theory \cite{PvsInfer2012,Asoodeh2014Note} consider random variables (rvs) $\Sen$ and $\Pub$, and assumes that an adversary infers statistical knowledge on $\Sen$ via $\Pub$.
By considering the correlation, the joint probability $p(\sen,\pub),\forall \sen, \pub$, between the private and public data, the problem is to design a private encoder/channel $p(\rel | \pub),\forall \pub,\rel$ that minimizes privacy leakage, but maintains some level of data utility.
This framework is also captured by the Markov chain $\Sen - \Pub - \Rel$, where $p(\rel | \pub)$ is called the \emph{privacy funnel} \cite{PF2014}. The design of the privacy funnel $p(\rel | \pub)$ is shown to be an opposite optimization problem to the \emph{information bottleneck}\footnote{For the utility function $\E_{\pub,\rel}[d(\pub,\rel)]$, the information bottleneck generalizes the rate-distortion problem in information theory \cite{IB2000}.} \cite{IB2000,IBAgglom1999}.
The idea of Algorithms~\ref{algo:AgloCluster2} and \ref{algo:AgloCluster1} in this paper is analogous to the agglomerative pairwise merge algorithms in \cite{IBAgglom1999,PF2014}.
Similar to $L_0(\Sen \rightarrow \Rel)$, privacy in information theory is measured in terms of the logarithm of the fraction between the prior and posterior statistical uncertainty on $\Sen$. For example, the average inference loss defined as $H(\Sen) - H(\Sen|\Rel)$, or the mutual information $I(\Sen ; \Rel)$, is extended to the worst case $H(\Sen) - \max_{\rel} H(\Sen | \rel)$ \cite{PvsInfer2012} and the $\alpha$-leakage \cite{Liao2019AlphaRobust,Liao2019Alpha}, where the latter is a tunable measure of the mutual information.

\subsection{Privacy Leakage $L_0$ and Stochastic Privacy Measures}

Let $\Sen$ be a uniformly distributed rv and consider the maximization of R\'{e}nyi divergence $D_\alpha(p(\sen|\rel) \| p(\sen)) = \frac{1}{\alpha - 1} \log \sum_{\rel} \frac{p^{\alpha}(\sen|\rel)}{p^{\alpha-1}(\sen)}$ in the order of $\alpha = 0$ over all $\rel \in \range{\Rel}$. We have
\begin{equation} \label{eq:MaxRenyi}
    \begin{aligned}
    \max_{\rel \in \range{\Rel}} D_0(p(\sen|\rel) \| p(\sen)) & = \max_{\rel \in \range{\Rel}} - \log \sum_{\sen \colon p(\sen | \rel) >0} p(s) \\
                                                              & = \max_{\rel \in \range{\Rel}} - \log \sum_{\sen \colon p(\sen | \rel) >0} \frac{1}{|\range{\Sen}|} \\
                                                              & = \max_{\rel \in \range{\Rel}} \log \frac{|\range{S}|}{|\range{S|\rel}|}\\
                                                              & = L_0(\Sen \rightarrow \Rel).
    \end{aligned}
\end{equation}
When the order $\alpha \rightarrow \infty$,
\begin{multline*}
    \max_{\rel \in \range{\Rel}} D_{\infty}(p(\sen|\rel) \| p(\sen))  = \max_{(\sen,\rel) \in \range{\Sen,\Rel}} \log \frac{p(\sen|\rel)}{p(\sen)} \\
     = D_{\infty}(p(\sen,\rel) \| p(\sen)p(\rel)) = \mathcal{L}^{r}(\Sen \rightarrow \Rel).
\end{multline*}
Here, $\mathcal{L}^{r}(\Sen \rightarrow \Rel)$ is called the \emph{maximal realizable leakage} in \cite[Definition~8]{Issa2016MaxL}.
It is also shown in \cite[Definition~6 and Theorem~3]{PvsInfer2012} that if
\begin{equation} \label{eq:LDP}
    \max_{(\sen,\rel) \in \range{\Sen,\Rel}} \Big| \log \frac{p(\sen|\rel)}{p(\sen)} \Big| = \epsilon,
\end{equation}
the randomization $p(\rel|\sen)$ is at least $2\epsilon$-locally differential private ($2\epsilon$-LDP)\footnote{A data release scheme $p(\pub|\sen)$ is $\epsilon$-LDP if $p(\pub | \sen)/p(\pub | \sen') \leq e^{\epsilon} , \forall \pub,\sen,\sen'$ \cite{LDP}. Note that LDP is equivalent to DP \cite{Dwork2011DP} without the constraint on the Hamming distance $d_H(\sen,\sen') \leq 1$. This definition is adopted in information theory, e.g., \cite{Sarwate2014LDP}.}.

In \cite[Theorem~1]{Issa2016MaxL}, the stochastic maximal leakage $\mathcal{L}(\Sen \rightarrow \Rel)$ is defined and shown to be equivalent to the Sibson mutual information $I_{\infty} (\Sen;\Rel)$, which is upper bounded as $\mathcal{L}(\Sen \rightarrow \Rel) \leq \mathcal{L}^{r}(\Sen \rightarrow \Rel)$ \cite[Section~VI-B]{Issa2016MaxL}.
Meanwhile, a more general measure, the maximal $\alpha$-leakage $\mathcal{L}_\alpha^{\text{max}}(\Sen \rightarrow \Rel)$, was proposed in~\cite[Definition~6]{Liao2019Alpha} based on the Arimoto mutual information, which reduces to the maximal leakage $\mathcal{L}_{\infty}^{\text{max}}(\Sen \rightarrow \Rel) = \mathcal{L}(\Sen \rightarrow \Rel)$ when the order $\alpha \rightarrow \infty$ \cite[Theorem~2]{Liao2019Alpha}.
It is also shown in \cite[Theorem~2]{Liao2019Alpha} that when $\alpha = 1$, the maximal $\alpha$-leakage $\mathcal{L}_{\alpha}^{\text{max}}(\Sen \rightarrow \Rel)$ reduces to Shannon mutual information $I(\Sen ; \Rel)$, which was used as the privacy measure, called \emph{synergistic disclosure capacity}, in \cite[Definition~2]{Rassouli2020}.
The mutual information $I(\Sen;\Rel)$ is equivalent to the $f$-information $I_f(\Sen;\Rel)$, e.g., used in \cite{Wang2018:Emp}, in the case of $f(t) = t \log t$.

\subsection{Maximin Information $I_*$ and G{\'a}cs-K{\"o}rner Common Information}

The maximin information $I_*(\Sen ; \Rel)$ is related to the G{\'a}cs-K{\"o}rner common information \cite{GacsKorner1973}:
$$K(\Sen ; \Rel)= - \sum_{\SEN \in \range{\Sen| \Rel}_*} p(\SEN) \log p(\SEN),
$$ where $p(\SEN) = \sum_{\sen \in \SEN, \rel \in \range{\Rel | \sen}} p(\sen, \rel)$ and $\range{\Sen,\Rel} = \Set{(\sen,\rel) \colon p(\sen,\rel) \neq 0}$.
Note that $I_* (\Sen ; \Rel) = \max_{p(\sen,\rel),\forall \sen,\rel} K(\Sen; \Rel)$ and therefore $K(\Sen; \Rel)$ denotes the number of bits in $\Sen$ that can be perfectly received by the adversary via the privacy funnel $p(\rel | \pub)$.
The sample complexity\footnote{The sample complexity derived in \cite[Proposition~2]{Xu2019Allerton} applies to learning the Hirschfeld-Gebelein-R\'{e}nyi (HGR) maximal correlation function, which determines $K(\Sen ; \Rel)$ \cite[Corollary~3, Remark 6]{GacsKorner2016arXiv} and therefore applies to $I_*(\Sen ; \Rel)$.} for empirically learning the value $K(\Sen ; \Rel)$ \cite[Proposition~2]{Xu2019Allerton} also applies to $I_*(\Sen ; \Rel)$.

Similar to $I_*(\Sen; \Rel) \leq L_0(\Sen \rightarrow \Rel)$, we have $K(\Sen ; \Rel) \leq I(\Sen ; \Rel)$ \cite{Salamatian2016GacsKorner,GacsKorner1973}. This means that, as long as $I(\Sen ; \Rel) > 0$, minimizing $I(\Sen;\Rel)$, or any other stronger privacy leakage, e.g., the maximal leakage \cite{Issa2016MaxL}, does not necessarily ensure $K(\Sen ; \Rel) = 0$. %
On the other hand, the case $I_*(\Sen; \Rel) > 0$ indicates that the LDP \cite{Dwork2011DP} is unattainable. This is because there exists an $\pub$ such that $\log (p(\pub | \sen)/p(\pub | \sen')) \rightarrow \infty$ for some pair of $\sen$ and $\sen'$.
%
%So, it is important to have $K(\Sen ; \Rel) = 0$ to ensure statistical indistinguishability. However, the problem of how to minimize the $K(\Sen; \Rel)$ is overlooked in the literature.

Note that, in stochastic information theory, \emph{perfect privacy} \cite{Rassouli2018PrefectPriv,Calmon2015PerfecPiv} refers to the independence $\Sen \perp \Rel$, i.e., $p(\sen | \rel ) = p(\sen), \forall \rel$. In this case, $I(\Sen; \Rel ) = 0$ so that $K(\Sen ; \Rel) = 0$, necessarily.
Although the condition $K(\Sen; \Rel) = 0$ alone does not guarantee perfect privacy\footnote{The interpretation is that even if the DP is attainable, the information leakage could be nonzero.}, the value of the G{\'a}cs-K{\"o}rner common information indicates an upper bound on the utility of the released data in the case of the perfect privacy (see~\cite{Asoodeh2014Note}).
The case $\Sen \perp \Rel$ corresponds to non-stochastic independence $\range{\Sen | \rel} = \range{\Sen},\forall \rel$, where $I_0(\Sen \rightarrow \Rel) = L_0(\Sen \rightarrow \Rel) = 0$ and $I_*(\Sen ; \Rel) = 0$. But, since $I_*(\Sen ; \Rel) \leq L_0(\Sen \rightarrow \Rel)$, $I_*(\Sen ; \Rel) = 0$ does not guarantee $\range{\Sen | \rel} = \range{\Sen},\forall \rel$, either. See also \cite[Fig.~5]{Nair2013}.
Therefore, the solution to problem~\eqref{eq:prob3} is not necessarily a data release scheme with perfect non-stochastic privacy.

  \begin{figure}[t]
        \centering
        \scalebox{0.6}{% This file was created by matlab2tikz v0.4.3.
% Copyright (c) 2008--2013, Nico Schlömer <nico.schloemer@gmail.com>
% All rights reserved.
%
% The latest updates can be retrieved from
%   http://www.mathworks.com/matlabcentral/fileexchange/22022-matlab2tikz
% where you can also make suggestions and rate matlab2tikz.
%
\begin{tikzpicture}

\begin{axis}[%
width=4in,
height=2in,
scale only axis,
separate axis lines,
xmin=0,
xmax=1,
xlabel={\large utility loss: $1 - \frac{U_1(\Pub;\Rel)}{U_1(\Pub;\Pub)}$ or $\frac{U_2(\Pub;\Rel)}{\hat{U}_2(\Pub;\Rel)}$},
ymin=0,
ymax=1,
ylabel={\large privacy leakage $I_*(\Sen;\Rel)/I_*(\Sen;\Pub)$},
grid=major,
legend style={at={(0.6,0.99)},anchor=north west,draw=black,fill=white,legend cell align=left}
]
\addplot [
color=blue,
line width=2.0pt,
mark size=4.0pt,
only marks,
mark=asterisk,
mark options={solid},
]
table[row sep=crcr]{
0 1\\
0.356207187108022 0.430676558073393\\
0.56457503405358 0\\
};
\addlegendentry{\large for problem \eqref{eq:prob1} when $ U_1 (\Sen;\Rel)$};

\addplot [
color=blue,
line width=2.0pt,
mark size=6.0pt,
only marks,
mark=pentagon,
mark options={solid},
]
table[row sep=crcr]{
0.56457503405358 0\\
};
\addlegendentry{\large solution determined in Example~\ref{ex:P1U1}};

\addplot [
color=red,
line width=2.0pt,
mark size=4.0pt,
only marks,
mark=x,
mark options={solid},
]
table[row sep=crcr]{
0 1\\
0.0256410256410256 0.861353116146786\\
0.0512820512820513 0.682606194485985\\
0.342051282051282 0.430676558073393\\
1 0\\
};
\addlegendentry{\large for problem \eqref{eq:prob1} when $U_2(\Sen;\Rel)$};

\addplot [
color=red,
line width=2.0pt,
mark size=6.0pt,
only marks,
mark=o,
mark options={solid},
]
table[row sep=crcr]{
0.342051282051282 0.430676558073393\\
};
\addlegendentry{\large solution determined in Example~\ref{ex:P1U2}};

\end{axis}
\end{tikzpicture}% 
}
   \caption{Pareto frontier for problem~\eqref{eq:prob1} for both utility functions $ U_1 (\Sen;\Rel) = I_0(\Pub \rightarrow \Rel)$ and $U_2(\Sen;\Rel) = - \max_{\pub,\rel \colon \rel \in \range{\Rel | \pub}} d(\pub,\rel)$. Note, both privacy and utility measures are normalized. For $ U_1 (\Sen;\Rel)$, the horizontal axis is $1 - \frac{U_1(\Pub;\Rel)}{U_1(\Pub;\Pub)} = 1-\frac{I_0(\Pub \rightarrow \Rel)}{H_0(\Pub)}$; For $ U_1 (\Sen;\Rel)$, the horizontal axis is $\frac{U_2(\Pub;\Rel)}{\hat{U}_2(\Pub;\Rel)}$, where $\hat{U}_2(\Pub;\Rel)$ is the minimum value of $U_2(\Pub;\Rel)$ over all points in the Pareto frontier.  }
   \label{fig:PvsU_toy}
 \end{figure}

\section{Experiments}
\label{sec:Exp}

In this section, we run experiments to show the performance of Algorithm~\ref{algo:AgloCluster2} and \ref{algo:AgloCluster1} proposed in Section~\ref{sec:Algo}. First, we show the Pareto frontier of Example~\ref{ex:main} to explain how to search the solution to problems~\eqref{eq:prob2}, \eqref{eq:prob1} and \eqref{eq:prob3}. Then, we present both the convergence performance and Pareto frontier of Algorithms~\ref{algo:AgloCluster2} and \ref{algo:AgloCluster1} when they are applied to the real-world dataset.

\subsection{Pareto Frontier}

The Lagrangian function $ \Lat (\range{\Rel | \Pub},\lambda) $ can be considered as the weighted-sum of two objective functions: the privacy leakage $I_*(\Sen;\Rel)$, or $L_0(\Sen \rightarrow \Rel)$, and the utility loss $- U_i(\Pub ; \Rel)$.
They form the PUT: minimizing one necessarily increases the other. In this case, the minimizer of the problem $ \min_{\range{\Rel | \Pub}}\Lat (\range{\Rel | \Pub},\lambda) $ for each value of $\lambda$ produces a \emph{Pareto optimal} pair $\Set{I_*(\Sen;\Rel), - U_i(\Pub ; \Rel)}$, or $\Set{L_0(\Sen \rightarrow \Rel), - U_i(\Pub ; \Rel)}$, indicating the smallest privacy leakage $I_*(\Sen;\Rel)$ or $L_0(\Sen \rightarrow \Rel)$ that can be attained by the design of the mapping $\range{\Rel | \Pub}$ to maintain certain level $\theta$ of $U_i(\Pub ;\ \Rel)$ and vice versa \cite{Hwang2012MultObj}.
The Pareto optimal pair for all value of $\lambda$ forms the \emph{Pareto frontier}.

Fig.~\ref{fig:PvsU_toy} shows the examples of the Pareto frontier for the problem~\eqref{eq:prob1}, where we can see the solutions obtained in Examples~\ref{ex:P1U1} and \ref{ex:P1U2}. %
Taking the solution in Example~\ref{ex:P1U1} for example. It corresponds to the pair: $I_*(\Sen ; \Rel) =0$ and $I_0(\Pub;\Rel) = \log 7/3$, which states that, to maintain the utility no less than $\log 7/3$, we can reduce the privacy leakage $I_*(\Sen; \Rel)$ to zero.
When applying the algorithms proposed in this paper to the real-world dataset, enumerate $\lambda$ to obtain the Pareto frontier. For given utility threshold $\theta$, search the Pareto frontier for the minimal privacy leakage and the corresponding solution $\range{\Rel|\Pub}$.

\begin{figure*}[tpb]
	\centering
    \subfigure[$\lambda = 2.3$ for $U_1(\Pub ; \Rel^{(t)})$; $\lambda = 2$ for $U_2(\Pub ; \Rel^{(t)})$]{\scalebox{0.6}{% This file was created by matlab2tikz v0.4.3.
% Copyright (c) 2008--2013, Nico Schlömer <nico.schloemer@gmail.com>
% All rights reserved.
%
% The latest updates can be retrieved from
%   http://www.mathworks.com/matlabcentral/fileexchange/22022-matlab2tikz
% where you can also make suggestions and rate matlab2tikz.
%
\begin{tikzpicture}

\begin{axis}[%
width=4.5in,
height=1.8in,
scale only axis,
xmin=0,
xmax=17,
xlabel={\large iteration index $t$},
ymin=-0.2,
ymax=1.8,
ylabel={\large $\Lat(\range{\Rel^{(k)} | \Pub}, \lambda)$},
]
\addplot [
color=blue,
solid,
line width = 1.5pt,
mark=asterisk,
]
table[row sep=crcr]{
0 1.64\\
1 0.28875\\
2 0.23\\
3 0.101333333333333\\
4 0.0560714285714286\\
5 0.0467857142857143\\
6 0.00625\\
7 0.00464285714285714\\
8 -0.00218137254901961\\
9 -0.00395604395604395\\
10 -0.00425714285714286\\
};
\addlegendentry{\large for problem \eqref{eq:prob2} when $U_1(\Sen;\Rel)$};

\addplot [
color=red,
solid,
line width = 1.5pt,
mark=o,
]
table[row sep=crcr]{
0 1.069375\\
1 0.876875\\
2 0.812708333333333\\
3 0.780625\\
4 0.761375\\
5 0.748541666666667\\
6 0.739375\\
7 0.7325\\
8 0.727152777777778\\
9 0.719375\\
10 0.613708333333333\\
11 0.607291666666667\\
12 0.6056875\\
13 0.604272058823529\\
14 0.597666666666667\\
15 0.597025\\
16 0.533125\\
17 0.53140625\\
};
\addlegendentry{\large for problem \eqref{eq:prob2} when $U_2(\Sen;\Rel)$};
\end{axis}
\end{tikzpicture}% 
}} \qquad
    \subfigure[$\lambda = 1.3$ for $U_1(\Pub ; \Rel^{(t)})$; $\lambda = 0.03$ for $U_2(\Pub ; \Rel^{(t)})$]{\scalebox{0.6}{% This file was created by matlab2tikz v0.4.3.
% Copyright (c) 2008--2013, Nico Schlömer <nico.schloemer@gmail.com>
% All rights reserved.
%
% The latest updates can be retrieved from
%   http://www.mathworks.com/matlabcentral/fileexchange/22022-matlab2tikz
% where you can also make suggestions and rate matlab2tikz.
%
\begin{tikzpicture}

\begin{axis}[%
width=4.5in,
height=1.8in,
scale only axis,
xmin=0,
xmax=80,
xlabel={\large iteration index $t$},
ymin=0.2,
ymax=1.4,
ylabel={\large $\Lat(\range{\Rel^{(k)} | \Pub}, \lambda)$},
]
\addplot [
color=blue,
line width = 1.5pt,
]
table[row sep=crcr]{
0 1.30052766953164\\
1 1.28335562756033\\
2 1.26626922676525\\
3 1.249268772799\\
4 1.23235457506574\\
5 1.21552694680047\\
6 1.19878620515068\\
7 1.18213267126054\\
8 1.16556667035758\\
9 1.14908853184214\\
10 1.13269858937957\\
11 1.11639718099536\\
12 1.10018464917329\\
13 1.08406134095671\\
14 1.06802760805315\\
15 1.05208380694232\\
16 1.0362302989877\\
17 1.02046745055186\\
18 1.00479563311562\\
19 0.989215223401362\\
20 0.97372660350048\\
21 0.958330161005327\\
22 0.943026289145768\\
23 0.927815386930585\\
24 0.912697859293942\\
25 0.897674117247161\\
26 0.882744578036054\\
27 0.867909665304072\\
28 0.85316980926157\\
29 0.838525446861457\\
30 0.823977021981582\\
31 0.80952498561416\\
32 0.795169796062627\\
33 0.780911919146267\\
34 0.766751828413045\\
35 0.752690005361068\\
36 0.738726939669112\\
37 0.724863129436732\\
38 0.711099081434451\\
39 0.697435311364594\\
40 0.683872344133364\\
41 0.670410714134771\\
42 0.657050965547132\\
43 0.64379365264282\\
44 0.630639340112084\\
45 0.617588603401741\\
46 0.604642029069672\\
47 0.591800215156044\\
48 0.579063771572333\\
49 0.566433320509246\\
50 0.553909496864751\\
51 0.541492948693517\\
52 0.529184337679162\\
53 0.51698433963084\\
54 0.504893645005807\\
55 0.492912959459743\\
56 0.481043004426785\\
57 0.469284517731358\\
58 0.457638254234105\\
59 0.446104986514419\\
60 0.434685505592289\\
61 0.423380621692451\\
62 0.412191165054099\\
63 0.401117986789732\\
64 0.390161959797061\\
65 0.379323979728302\\
66 0.368604966021588\\
67 0.358005862999771\\
68 0.347527641042411\\
69 0.337171297837364\\
70 0.326937859719127\\
71 0.316828383101852\\
72 0.306843956015889\\
73 0.296985699757728\\
74 0.287254770664398\\
75 0.277652362024747\\
76 0.268179706141553\\
77 0.258838076560242\\
};
\addlegendentry{\large for problem \eqref{eq:prob1} when $U_1(\Sen;\Rel)$};

\addplot [
color=red,
line width = 1.5pt,
]
table[row sep=crcr]{
0 1.28163402452584\\
1 1.26846198255453\\
2 1.25537558175945\\
3 1.24237512779319\\
4 1.22946093005993\\
5 1.21663330179466\\
6 1.20389256014488\\
7 1.19123902625474\\
8 1.17867302535178\\
9 1.16619488683633\\
10 1.15380494437376\\
11 1.14150353598956\\
12 1.12929100416749\\
13 1.11716769595091\\
14 1.10513396304734\\
15 1.09319016193651\\
16 1.08133665398189\\
17 1.06957380554605\\
18 1.05790198810981\\
19 1.04632157839556\\
20 1.03483295849467\\
21 1.02343651599952\\
22 1.01213264413996\\
23 1.00092174192478\\
24 0.989804214288135\\
25 0.978780472241354\\
26 0.963850933030247\\
27 0.949016020298265\\
28 0.934276164255763\\
29 0.91963180185565\\
30 0.905083376975775\\
31 0.890631340608353\\
32 0.876276151056821\\
33 0.86201827414046\\
34 0.847858183407238\\
35 0.833796360355261\\
36 0.819833294663306\\
37 0.805969484430925\\
38 0.792205436428644\\
39 0.778541666358788\\
40 0.764978699127557\\
41 0.751517069128964\\
42 0.738157320541325\\
43 0.724900007637013\\
44 0.711745695106277\\
45 0.698694958395935\\
46 0.685748384063866\\
47 0.672906570150237\\
48 0.660170126566526\\
49 0.647539675503439\\
50 0.635015851858945\\
51 0.62259930368771\\
52 0.610290692673355\\
53 0.598090694625033\\
54 0.586\\
55 0.574019314453937\\
56 0.562149359420978\\
57 0.550390872725551\\
58 0.538744609228298\\
59 0.527211341508612\\
60 0.515791860586482\\
61 0.504486976686644\\
62 0.493297520048292\\
63 0.482224341783925\\
64 0.471268314791255\\
65 0.460430334722496\\
66 0.449711321015781\\
67 0.439112217993964\\
68 0.428633996036604\\
69 0.418277652831558\\
70 0.40804421471332\\
71 0.397934738096045\\
72 0.387950311010082\\
73 0.378092054751921\\
74 0.368361125658592\\
75 0.35875871701894\\
76 0.349286061135746\\
77 0.339944431554435\\
78 0.330735145476215\\
79 0.321659566375852\\
80 0.321290535418537\\
};
\addlegendentry{\large for problem \eqref{eq:prob1} when $U_2(\Sen;\Rel)$};

\end{axis}
\end{tikzpicture}% 
}}
    %\scalebox{0.6}{\input{figures/Converge_Istat_heartdisease.tex}}
	\caption{The convergence performance in terms of $\Lat(\range{\Rel^{(t)} | \Pub},\lambda) $ versus the iteration index $t$ when Algorithms~\ref{algo:AgloCluster2} and \ref{algo:AgloCluster1} are applied to the Hungarian heart disease dataset in \cite{UCI2007} for solving problems~\eqref{eq:prob2} and \eqref{eq:prob1}, respectively. Here, $U_1(\Pub ; \Rel^{(t)}) = I_0(\Sen \rightarrow \Rel^{(t)})$ and $U_2(\Pub ; \Rel^{(t)}) = -\max_{\pub,\rel \colon \rel \in \range{\Rel^{(t)} | \pub}} d(\pub,\rel)$. }
	\label{fig:Convergence_Hungarian_Istar}
\end{figure*}

  \begin{figure*}[th]
        \centering
        \subfigure{\scalebox{0.6}{% This file was created by matlab2tikz v0.4.3.
% Copyright (c) 2008--2013, Nico Schlömer <nico.schloemer@gmail.com>
% All rights reserved.
%
% The latest updates can be retrieved from
%   http://www.mathworks.com/matlabcentral/fileexchange/22022-matlab2tikz
% where you can also make suggestions and rate matlab2tikz.
%
\begin{tikzpicture}

\begin{axis}[%
width=4in,
height=2in,
scale only axis,
separate axis lines,
xmin=0,
xmax=1,
xlabel={\large utility loss: $1 - \frac{U_1(\Pub;\Rel)}{U_1(\Pub;\Pub)}$ or $\frac{U_2(\Pub;\Rel)}{\hat{U}_2(\Pub;\Rel)}$},
ymin=0,
ymax=1,
ylabel={{\large privacy leakage} $L_0(\Sen \rightarrow \Rel)/L_0(\Sen \rightarrow \Pub)$},
grid=major,
legend style={at={(0.35,0.99)},anchor=north west,draw=black,fill=white,legend cell align=left}
]
\addplot [
color=blue,
line width=2.0pt,
mark size=4.0pt,
only marks,
mark=asterisk,
mark options={solid},
]
table[row sep=crcr]{
0.96 0.00371057513914657\\
0.952380952380952 0.004995004995005\\
0.941176470588235 0.00757575757575758\\
0.928571428571429 0.012987012987013\\
0.916666666666667 0.0155844155844156\\
0.857142857142857 0.0454545454545455\\
0.833333333333333 0.0538033395176252\\
0.8 0.0822510822510822\\
0.666666666666667 0.168831168831169\\
0.5 0.224025974025974\\
0 1\\
};
\addlegendentry{\large Algorithm~\ref{algo:AgloCluster2} for problem \eqref{eq:prob2} when $ U_1 (\Sen;\Rel)$};

\addplot [
color=red,
line width=2.0pt,
mark size=4.0pt,
only marks,
mark=o,
mark options={solid},
]
table[row sep=crcr]{
0.241095890410959 0.34375\\
0.15013698630137 0.44\\
0 1\\
};
\addlegendentry{\large Algorithm~\ref{algo:AgloCluster2} for problem \eqref{eq:prob2} when $U_2(\Sen;\Rel)$};

\addplot [
color=green,
line width=2.0pt,
mark size=2.8pt,
only marks,
mark=pentagon,
mark size=4.0pt,
mark options={solid},
]
table[row sep=crcr]{
0.7 0.168831168831169\\
};
\addlegendentry{\large Method \cite[Section~3.1]{Sweeney2002} attains $5$-anonymity};

\end{axis}
\end{tikzpicture}% 
}} \quad
        \subfigure{\scalebox{0.6}{% This file was created by matlab2tikz v0.4.3.
% Copyright (c) 2008--2013, Nico Schlömer <nico.schloemer@gmail.com>
% All rights reserved.
%
% The latest updates can be retrieved from
%   http://www.mathworks.com/matlabcentral/fileexchange/22022-matlab2tikz
% where you can also make suggestions and rate matlab2tikz.
%
\begin{tikzpicture}

\begin{axis}[%
width=4in,
height=2in,
scale only axis,
separate axis lines,
xmin=0,
xmax=1,
xlabel={\large utility loss: $1 - \frac{U_1(\Pub;\Rel)}{U_1(\Pub;\Pub)}$ or $\frac{U_2(\Pub;\Rel)}{\hat{U}_2(\Pub;\Rel)}$},
ymin=0,
ymax=1,
ylabel={{\large privacy leakage} $I_*(\Sen;\Rel)/I_*(\Sen;\Pub)$},
grid=major,
legend style={at={(0.35,0.99)},anchor=north west,draw=black,fill=white,legend cell align=left}
]

\addplot [
color=blue,
line width=2.0pt,
mark size=4.0pt,
only marks,
mark=asterisk,
mark options={solid},
]
table[row sep=crcr]{
0 1\\
0.137612408786178 0.338983050847458\\
0.218110507560002 0.127118644067797\\
0.275224817572355 0.0254237288135593\\
0.31952611816575 0.00847457627118644\\
};
\addlegendentry{\large Algorithm~\ref{algo:AgloCluster1} for problem \eqref{eq:prob1} when $ U_1 (\Sen;\Rel)$};

\addplot [
color=red,
line width=2.0pt,
mark size=2.5pt,
only marks,
mark=o,
mark options={solid},
]
table[row sep=crcr]{
0.00606720597386434 1\\
0.0364032358431861 0.957264957264957\\
0.0667392657125078 0.914529914529915\\
0.0970752955818295 0.871794871794872\\
0.115276913503423 0.769230769230769\\
0.121344119477287 0.752136752136752\\
0.127411325451151 0.700854700854701\\
0.133478531425016 0.598290598290598\\
0.13954573739888 0.564102564102564\\
0.145612943372744 0.427350427350427\\
0.151680149346609 0.333333333333333\\
0.164681305004889 0.282051282051282\\
0.172048626544582 0.273504273504274\\
0.186060983198506 0.264957264957265\\
0.20059699751089 0.196581196581197\\
0.311197106409459 0.188034188034188\\
0.543621655258245 0.128205128205128\\
0.623843600912674 0.111111111111111\\
0.737106043413009 0.094017094017094\\
0.769206151657925 0.0598290598290598\\
0.85301636421736 0.0341880341880342\\
0.967091710834066 0.0256410256410256\\
1 0.00854700854700855\\
};
\addlegendentry{\large Algorithm~\ref{algo:AgloCluster1} for problem \eqref{eq:prob1} when $U_2(\Sen;\Rel)$};

\addplot [
color=black,
line width=2.0pt,
mark size=2.8pt,
only marks,
mark=square,
mark options={solid},
]
table[row sep=crcr]{
0.132568450528936 0.853846153846154\\
};
\addlegendentry{\large \cite[Algorithm~1]{Ding2019ITW} for problem \eqref{eq:prob1} when $U_2(\Sen;\Rel)$};
\end{axis}
\end{tikzpicture}% 
}}
   \caption{Pareto frontiers for the problems~\eqref{eq:prob2} and \eqref{eq:prob1} for both utility functions $ U_1 (\Sen;\Rel) = I_0(\Pub \rightarrow \Rel)$ and $U_2(\Sen;\Rel) = - \max_{\pub,\rel \colon \rel \in \range{\Rel | \pub}} d(\pub,\rel)$, obtained by applying Algorithms~\ref{algo:AgloCluster2} and \ref{algo:AgloCluster1} to the Hungarian heart disease dataset in \cite{UCI2007}. They are compared to the solution search by the generalization-and-suppression algorithm for attaining the $5$-anonymity in \cite[Section~3.1]{Sweeney2002} and the submodularity-based agglomerative clustering algorithm in \cite[Algorithm~1]{Ding2019ITW}.}
   \label{fig:PvsU_Hungarian_Istar}
 \end{figure*}

\subsection{Hungarian Heart Disease Dataset}
\label{sec:Hungarian}

This experiment is based on the heart disease data set created by the Hungarian Institute of Cardiology, Budapest, in the UCI machine learning repository \cite{UCI2007}. It records $293$ patients' data of 76 attributes for the purpose of identifying the presence of heart disease.
We extract the column `age' as the sensitive data $\Sen$ and `serum cholesterol (mg/dl)' as the public data $\Pub$. The marginal range $\range{\Sen}$ contains $38$ distinct values of age and $\range{\Pub}$ contains $154$ cholesterol levels.
The joint range $\range{\Sen,\Pub}$ contains $281$ distinct values of $(\sen,\pub)$, among which $269$ values appear only once in the whole dataset.
Note that this is a typical example that the size of released dataset may not be large enough to reveal the true statistics of the data.
In this case, each data record often releases a new value of $(\sen,\pub)$ and the adversary only infers the supports or cardinalities of the alphabets $\mathcal{\Sen}$ and $\mathcal{\Pub}$, not the probability or data statistics.
Therefore, the variables $\Sen$ and $\Pub$ can be treated as uncertain variables and the information can be directly obtained from the ranges of $\Sen$ and $\Pub$ by the non-stochastic information measures in Section~\ref{sec:PrivMeasure}.

Fig.~\ref{fig:Convergence_Hungarian_Istar} shows that Algorithms~\ref{algo:AgloCluster2} and \ref{algo:AgloCluster1} converge to the locally optimal solutions to problems \eqref{eq:prob2} and \eqref{eq:prob1}, respectively.
Fig.~\ref{fig:PvsU_Hungarian_Istar} shows the Pareto frontier searched by Algorithms~\ref{algo:AgloCluster2} and \ref{algo:AgloCluster1} by enumerating the value of Lagrangian multiplier $\lambda$.

Because of the equivalence of $k$-anonymity and $L_0$ in Lemma~\ref{lemma2}, we also run the generalization-and-suppression method proposed~\cite[Section~3.1]{Sweeney2002} for attaining the $k$-anonymity for $k = 5$. The idea is to search for all $\pub$ associated with less than $5$ values of $\sen$ and merge them to a new but less specific value, which is exactly the same idea as the agglomerative clustering algorithms proposed in this paper.
The only difference is that the generalization-and-suppression method does not guarantee a certain level of utility, which means it may sacrifice too much data utility to attain the $k$-anonymity.
As shown in Fig.~\ref{fig:PvsU_Hungarian_Istar}, the solution determined by Algorithm~\ref{algo:AgloCluster2} outperforms the generalization-and-suppression method proposed in~\cite[Section~3.1]{Sweeney2002} in that it incurs a lower utility loss, or maintains a higher level of usefulness in the released dataset.\footnote{It can be seen by comparing to the blue star searched by Algorithm~\ref{algo:AgloCluster2} on the left-hand side of the green pentagon determined by the generalization-and-suppression method proposed in~\cite[Section~3.1]{Sweeney2002}.}

For problem~\eqref{eq:prob1}, we run the submodularity-based agglomerative clustering algorithm proposed in \cite[Algorithm~1]{Ding2019ITW} and compare with Algorithm~\ref{algo:AgloCluster1} in the Pareto frontier in Fig.~\ref{fig:PvsU_Hungarian_Istar}. It can be seen that Algorithm~\ref{algo:AgloCluster1} is able to search a solution that \emph{Pareto dominates}~\cite[Algorithm~1]{Ding2019ITW}: One can find several points determined by Algorithm~\ref{algo:AgloCluster1} with both privacy leakage and utility loss strictly less than the solution searched by~\cite[Algorithm~1]{Ding2019ITW}. This is because the problem under concern is not submodular\footnote{The validity and efficiency of the agglomerative clustering algorithm in \cite[Algorithm~1]{Ding2019ITW} is based on the submodularity of the Lagrangian function.
However, in this paper, the Lagrangian function $\Lat (\range{\Rel^{(t)} | \Pub},\lambda)$ for both privacy measures $I_*$ and $L_0$ does not exhibit submodularity or supermodularity. One can prove by counter-examples based on the definition of submodularity in \cite{Bach2010SFMtut,Fujishige2005}.}
and therefore submodularity-based algorithms may not be superior to greedy ones.

\section{Conclusion}

We studied the problem of how to quantize the public data $\Pub$ into $\Rel$ to minimize the non-stochastic information leakage of the private data $\Sen$ but guarantee a certain level of the data utility.
For the privacy measures $L_0(\Sen \rightarrow \Rel)$ and $I_*(\Sen ; \Pub)$, two agglomerative clustering algorithms were proposed, respectively, both of which generate a solution $\range{\Rel | \Pub}$ by recursively merging two elements in $\range{\Pub}$ that strictly reduces the Lagrangian function.
We showed that $L_0(\Sen \rightarrow \Rel)$ measures the maximum (worst-case) posterior range/uncertainty reduction on $\Sen$ at the adversary side and $I_*(\Sen ; \Rel)$ denotes how distinguishable $\Sen$ is by observing $\Rel$.
We then applied the clustering algorithm for minimizing $I_*(\Sen ; \Pub)$ to search for a $\Rel$ that guarantees non-stochastic indistinguishability $I_*(\Sen; \Rel) = 0$ but minimizes $L_0(\Sen \rightarrow \Rel)$ subject to some utility constraint.
It is shown that the value of $I_*(\Sen;\Rel)$ is equal to the maximum number of disconnected subgraphs in the confusability graph, which can be determined by the min-cut or optimal network attack algorithms. This gives a visualization of
the clustering algorithm for minimizing $I_*(\Sen; \Pub)$ (Algorithm~\ref{algo:AgloCluster1}) as a subgraph merging process.

There are some aspects in this paper that can be further explored. First, the proposed greedy clustering algorithms only converge to a local optimum and have the complexity $O(|\range{\Pub}|^3)$. It is worth discussing how to improve the accuracy and efficiency.
Second, based on Section~\ref{sec:ToITPriv}, we can combine the results in stochastic and non-stochastic information-theoretic privacy in real applications, e.g., adopt non-stochastic measure $L_0(\Sen \rightarrow \Rel)$ at the beginning of the data release and then switch\footnote{For example, change to a stochastic measure when the amount of the released data is enough (say, 5000 tabular records). In this case, it is also crucial to determine the optimal switching time. }
to stochastic measure $\max_{\rel \in \range{\Rel}} D_\alpha(p(\sen|\rel) \| p(\sen))$, which is done by increasing $\alpha$ from $0$ based on \eqref{eq:MaxRenyi}, or start with $I_*(\Sen;\Rel)$ and switch to $K(\Sen;\Rel)$.
In addition, the non-stochastic privacy measures can also be utilized to design stochastic privacy-preserving mechanisms: determining the optimal transition probability as well as the optimal alphabet size.\footnote{Most stochastic privacy-preserving solutions only determine the optimal transition probability $p(\rel | \pub)$ for fixed size of $\range{\Rel}$, e.g., the variational method in~\cite{IB2000} for information bottleneck, the Blahut–Arimoto algorithm in~\cite{Blahut1972} for rate-distortion.
An iterative approach can be proposed by combining these methods with Algorithms~\ref{algo:AgloCluster2} and \ref{algo:AgloCluster1} to first determine the optimal alphabet $\range{\Rel}$, then the optimal transition probability $p(\rel | \pub)$ (see also the deterministic annealing method in~\cite{DA1998}).
A related study can be found in~\cite{Shanir2010GIB}, which shows how, in information bottleneck, the error of empirical mutual information is upper bounded by the alphabet size.}
Third, the same as the \GK common information, the maximin information is zero in most cases. Therefore, it is of interest to consider other non-stochastic measure of the common information, e.g., to propose a non-stochastic version of the Wyner's common information in \cite{Wyner1975}, and explore its role in perfect privacy.

%it is of interest to further explore the role of the maximin information and the G{\'a}cs-K{\"o}rner common information, as well as other common information measures \cite{Asoodeh2014Note}, e.g., the Wyner's common information \cite{Wyner1975}, in perfect privacy and distinguishability in both stochastic and non-stochastic information-theoretic privacy.
%
%In addition, the relationship between the maximin information $I_*$ and other probabilistic measures, e.g., the maximum correlation coefficient and principal inertia component \cite{Calmon2017PIC}, is also worth studying .

% or


\appendices

\section{Proof of Proposition~\ref{prop:1}} \label{proof:prop:1}

For the finest $\range{\Sen | \Pub}$-overlap partition $\Pat_{\range{\Sen|\Pub}}^*$ and the maximin information $I_*(\Sen ; \Pub) = \log |\Pat_{\range{\Sen | \Pub}}^*|$, we have the Hartley entropy of $\Sen$ being

\begin{equation} \label{eq:ProofBound}
    \begin{aligned}
        H_0( \Sen ) & = \log |\range{\Sen}| \\
                    & = \log \sum_{\SEN \in \Pat_{\range{\Sen | \Pub}}^*} |\SEN| \\
                    & = I_*(\Sen ; \Pub) + \log \Big( \frac{1}{|\Pat_{\range{\Sen | \Pub}}^*|} \sum_{\SEN \in \Pat_{\range{\Sen | \Pub}}^*} |\SEN| \Big) \\
                    & \geq I_*(\Sen ; \Pub) + \log  \min_{\SEN \in \Pat_{\range{\Sen | \Pub}}^*} |\SEN| \\
                    & \geq I_*(\Sen ; \Pub) + \log  \min_{\pub \in \range{\Pub}} |\range{\Sen | \pub}|.
    \end{aligned}
\end{equation}
This proves $I_*(\Sen ; \Pub) \leq \log \frac{|\range{\Sen}|}{ \min_{\pub \in \range{\Pub}} |\range{\Sen | \pub}|} = \max_{\pub \in \range{\Pub}}  \log \frac{|\range{\Sen}|}{ |\range{\Sen | \pub}|} = L_0(\Sen \rightarrow \Pub)$.
Note that the inequality $K(\Sen ; \Pub) \leq I(\Sen ; \Pub)$ \cite{Salamatian2016GacsKorner,GacsKorner1973} (see Section~\ref{sec:ToITPriv}) describing the relation between the
G{\'a}cs-K{\"o}rner common information $K(\Sen ; \Pub)$ and Shannon mutual information $I(\Sen ; \Pub)$ in stochastic information theory can be proved in the same way as \eqref{eq:ProofBound} based on the Jensen's inequality. \hfill \IEEEQED

%\farhad{Let $I_*(\Pub;\Sen)=m$ and  $\range{\Pub|\Sen}_\star=\{P_1,\dots,P_{2^m}\}$. Each $P_i$ is non-empty. Therefore, there exists at least one $x$ such that  $x\in P_i$. Note that $x$ must also belong to $\range{\Pub|s}$ for any $s\in\range{\Sen|x}$. We prove that $\range{\Pub|s}\subseteq P_i$.  Assume that this not the case. Therefore, there exist an element of $x'\in\range{\Pub|s}$, distinct from $x$, that belongs to another $P_j$, $j\neq i$, because $\{P_1,\dots,P_{2^m}\}$ covers $\range{\Pub}$. We know that $P_i$ and $P_j$ are $\range{\Pub|\Sen}$-overlap isolated by the definition of partition $\range{\Pub|\Sen}_\star$. On the other hand, we evidently have $x\leftrightsquigarrow x'$ (by the definition of $\range{\Pub|\Sen}$-overlap connectedness). This is a contradiction and thus $\range{\Pub|s}$ must be a subset of $P_i$. This results in $|\range{\Pub|s}|\leq |P_i|$ and hence
%\begin{align}
%\min_{y\in\range{\Sen}}|\range{\Pub|s}|\leq P_i, \quad \forall i\in\{1,\dots,2^m\}. \label{eqn:proof:1}
%\end{align}
%On the other hand, $\bigcup_{i=1}^{2^m}P_i=\range{\Pub}$ because $\{P_1,\dots,P_{2^m}\}$ is a partition for $\range{\Pub}$. Because of the non-overlapping nature of the sets $\{P_1,\dots,P_{2^m}\}$, we get
%\begin{align}
%\sum_{i=1}^{2^m} |P_i|=|\range{\Pub}|.\label{eqn:proof:2}
%\end{align}
%Combining~\eqref{eqn:proof:1} and~\eqref{eqn:proof:2} results in $2^m\min_{s\in\range{\Sen}}|\range{\Pub|s}|\leq |\range{\Pub}|$. This concludes the proof.
%}

\section{Proof of Lemma~\ref{lemma:Graph}} \label{app:lemma:Graph}
    For $\PUB \subseteq \range{\Pub}$, denote $\range{\Sen | \PUB } = \bigcup_{\pub \in \PUB} \range{\Sen | \Pub = \pub}$. Since there is no edge connecting any distinct subgraphs $\mathcal{\Pub},\mathcal{\Pub}' \in \Pat_{\range{\Pub}}$, i.e., $ \range{\Sen | \PUB} \cap \range{\Sen | \PUB'} = \emptyset$, and
   $\bigcup_{\mathcal{\Pub} \in \Pat_{\range{\Pub}}} \range{\Sen | \PUB} = \bigcup_{\pub \in \range{\Pub}} \range{\Sen | \Pub = \pub} = \range{\Sen}. $
   So, $\Set{\range{\Sen | \PUB} : \mathcal{\Pub} \in \Pat_{G_{\Pub}}}$ for every $\Pat_{G_{\Pub}}$ is a partition of $\range{\Sen}$.
    For any two points $\pub \in \mathcal{\Pub}$ and $\pub' \in \mathcal{\Pub}'$ such that $\mathcal{\Pub},\mathcal{\Pub}'$ are two distinct subgraphs in $\Pat_{G_{\Pub}}$, there does not exist a path, denoted by an ordered sequence $(\pub_1,\dotsc,\pub_n)$, such that $\pub_1 = \pub$ and $\pub_n = \pub'$ and $(\pub_i,\pub_{i+1}) \in \Edge_{\range{\Sen | \Pub}}$, i.e., $\range{\Sen | \Pub = \pub_i} \cap \range{\Sen | \Pub = \pub_{i+1}} \neq \emptyset$. This means any $\sen \in \range{\Sen | \Pub = \pub} \subseteq \range{\Sen | \PUB}$ and $\sen' \in \range{\Sen | \Pub = \pub'} \subseteq \range{\Sen | \PUB'}$ are overlap isolated. Therefore, $\Pat_{G_{\Pub}}$ is a $\range{\Sen | \Pub}$-overlap isolated partition
    In the finest decomposition $\Pat_{G_{\Pub}}^*$, for each $\mathcal{X} \in \Pat_{G_{\Pub}}^*$, every pair $\pub, \pub' \in \mathcal{X}$ are connected, i.e., there exists a path $(\pub_1 = \pub,\dotsc,\pub_n = \pub')$ such that $(\pub_i,\pub_{i+1}) \in \Edge_{\range{\Sen | \Pub}}$ or $\range{\Sen | \Pub = \pub_i} \cap \range{\Sen | \Pub = \pub_{i+1}} \neq \emptyset$. Then, we have $\sen \Link \sen'$ for any two $\sen , \sen' \in \range{\Sen | \PUB}$. So, $\Set{\range{\Sen | \PUB} : \mathcal{\Pub} \in \Pat_{G_{\Pub}}^*}$ is $\range{\Sen | \Pub}$-overlap partition and also the finest one since it can not be further decomposed. \hfill\IEEEQED

%% use section* for acknowledgment
%\section*{Acknowledgment}

\ifCLASSOPTIONcaptionsoff
  \newpage
\fi

\bibliographystyle{IEEEtran}
\bibliography{NonStochastic}

\begin{IEEEbiography}[{\includegraphics[width=1in,height=1.25in,clip,keepaspectratio]{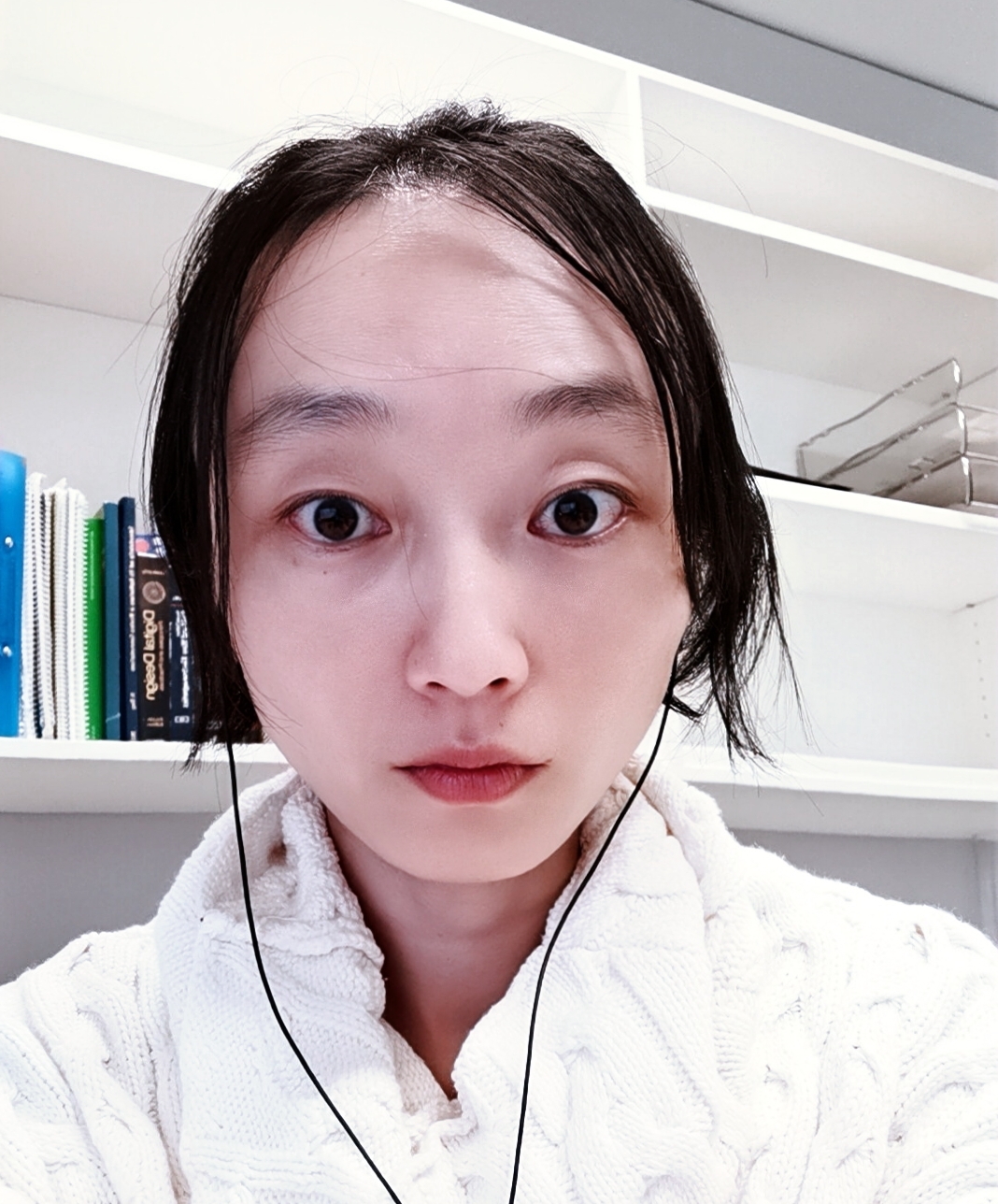}}]{Ni Ding}
received the B.C.A. degree from the Shanghai Second Polytechnic University, China, the B.E. degree in Telecommunications (1st class honours) from the University of New South Wales, Australia, and the PhD degree from the Australian National University, Australia, in 2005, 2012 and 2017, respectively. She was a postdoctoral fellow at Data 61, The Commonwealth Scientific and Industrial Research Organisation (CSIRO), Australia, from 2017 to 2019. She is now a Doreen Thomas Postdoctoral Fellow at the School of Computing and Information Systems, University of Melbourne.
Her research interests generally include optimizations in information theory, wireless communications, signal processing and machine learning. She is currently interested in data privacy, discrete and combinatorial optimization problems raised in discrete event control in cross-layer adaptive modulation, source coding and game theory (in particular, the games with strong structures, e.g., supermodular and convex games).
\end{IEEEbiography}

\begin{IEEEbiography}[{\includegraphics[width=1in,height=1.25in,clip,keepaspectratio]{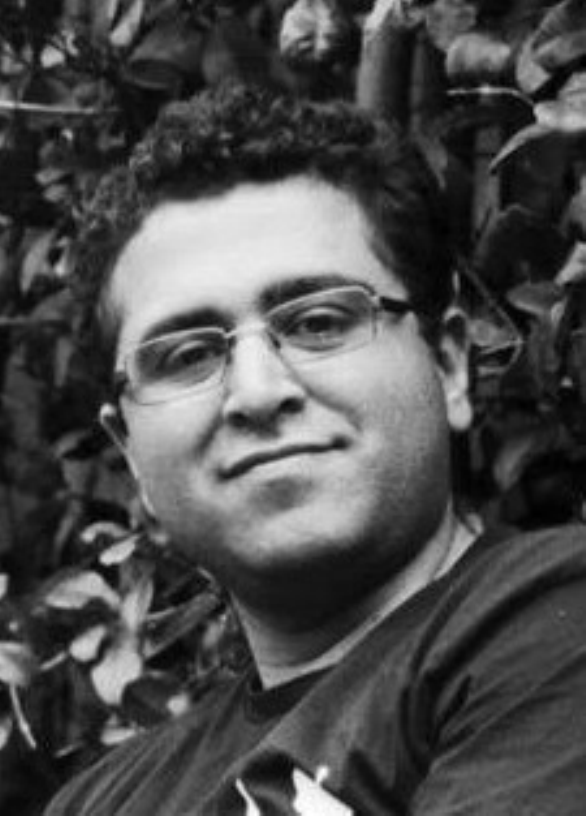}}]%
{Farhad Farokhi} (Senior Member, IEEE) received the Ph.D. degree from the KTH Royal Institute of Technology in 2014. He is currently a Lecturer (Assistant Professor) with the Department of Electrical and Electronic Engineering, The University of Melbourne. Prior to that, he was a Research Scientist with the Information Security and Privacy Group at CSIRO's Data61, a Research Fellow at The University of Melbourne, and a Post-Doctoral Fellow with the KTH Royal Institute of Technology. During his Ph.D. studies, he was a Visiting Researcher with the
University of California at Berkeley and the University of Illinois at Urbana-Champaign. He was a recipient of the VESKI Victoria Fellowship from the Victorian State Government, and the McKenzie Fellowship and the 2015 Early Career Researcher Award from The University of Melbourne. He was a Finalist in the 2014 European Embedded Control Institute (EECI) Ph.D. Award. He has been part of numerous projects on data privacy and cyber-security funded by the Defence Science and Technology Group (DSTG), the Department of the Prime Minister and Cabinet (PMC), the Department of Environment and Energy (DEE), and CSIRO, Australia.
\end{IEEEbiography}

%\end{IEEEbiography}
%% insert where needed to balance the two columns on the last page with
%%\newpage

%\vfill


\end{document}